\documentclass[draft]{agujournal2019}
\usepackage{url} 
\usepackage{lineno}
\usepackage{flafter}
\usepackage[inline]{trackchanges} 
\usepackage{soul}
\usepackage{appendix}
\usepackage{array}
\usepackage{multirow}
\usepackage{amsfonts}
\usepackage{amsmath}
\usepackage{xcolor}
\usepackage{nameref}
\usepackage{amssymb}
\usepackage{xr}
\usepackage{todonotes}

\let\oldcite\cite
\let\oldciteA\citeA
\renewcommand{\cite}[2][]{%
    \textcolor{blue}{\oldcite[#1]{#2}}%
}

\renewcommand{\citeA}[2][]{%
    \textcolor{blue}{\oldciteA[#1]{#2}}%
}

\DeclareRobustCommand{\colorednameref}[1]{%
    \textcolor{blue}{\nameref{#1}}%
}

\newcommand{\Mw}{M_\mathrm{w}}

\draftfalse

\begin{document}

\title{Ground Motion Characteristics of Cascading Earthquakes in a Multiscale Fracture Network}

\authors{Kadek Hendrawan Palgunadi\affil{1,\dagger}, Alice-Agnes Gabriel\affil{2,4}, Dmitry Igor Garagash\affil{3}, Thomas Ulrich\affil{4}, Nico Schliwa\affil{4}, Paul Martin Mai\affil{1}}

\affiliation{1}{Physical Science and Engineering, King Abdullah University of Science and Technology, Thuwal, Saudi Arabia}
\affiliation{2}{Institute of Geophysics and Planetary Physics, Scripps Institution of Oceanography, University of California, San Diego, CA, USA}
\affiliation{3}{Dalhousie University, Department Civil Resource Engineering, Halifax, Canada}
\affiliation{4}{Department of Earth and Environmental Sciences, Geophysics, Ludwig-Maximilians-Universit\"{a}t M\"{u}nchen, Munich, Germany}
\affiliation{\dagger}{now at Swiss Seismological Service (SED), ETH Zürich, Switzerland; Department of Geophysical Engineering, Institut Teknologi Sepuluh Nopember, Indonesia.}

\correspondingauthor{Authors}{kadek.palgunadi@its.ac.id}

\vspace{15mm}
{\textbf{
This manuscript is an arXiv preprint that has been submitted to a peer-reviewed journal and has not yet undergone peer review.
}}

\begin{abstract}
Fault zones exhibit geometrical complexity and are often surrounded by multiscale fracture networks within their damage zones, potentially influencing rupture dynamics and near-field ground motions. In this study, we investigate the ground-motion characteristics of cascading ruptures across damage zone fracture networks of moderate-sized earthquakes ($\Mw$ 5.5–6.0) using high-resolution 3D dynamic rupture simulations. Our models feature a listric fault surrounded by more than 800 fractures, emulating a major fault and its associated damage zone. 
We analyze three cases: a cascading rupture propagating within the fracture network ($\Mw$ 5.5), a non-cascading main-fault rupture with off-fault fracture slip ($\Mw$ 6.0), and a main-fault rupture without a fracture network ($\Mw$ 6.0).
Cascading ruptures within the fracture network produce distinct ground-motion signatures with enriched high-frequency content, arising from simultaneous slip of multiple fractures and parts of the main fault, resembling source coda-wave-like signatures. 
This case shows elevated near-field characteristic frequency ($f_c$) and stress drop, approximately an order of magnitude higher than the estimation directly on the fault of the dynamic rupture simulation.
The inferred $f_c$ of the modeled vertical ground-motion components reflects the complexity of the radiation pattern and rupture directivity of fracture-network cascading earthquakes. We show that this is consistent with observations of strong azimuthal dependence of corner frequency in the 2009-2016 Central Apennines, Italy, earthquake sequence. 
Simulated ground motions from fracture-network cascading ruptures also show pronounced azimuthal variations in peak ground acceleration (PGA), peak ground velocity, and pseudo-spectral acceleration, with average PGA nearly double that of the non-cascading cases. 
Cascading ruptures radiate high-frequency seismic energy, yield non-typical ground motion characteristics including coda-wave-like signatures, and may result in a significantly higher seismologically inferred stress drop and PGA. Such outcomes emphasize the critical role of fault-zone complexity in affecting rupture dynamics and seismic radiation and have important implications for physics-based seismic hazard assessment.
\end{abstract}

\begin{keypoints}
\item We analyze simulated ground motion from high-resolution 3D dynamic rupture of cascading and non-cascading earthquakes.
\item The cascading rupture radiates high-frequency seismic energy and yields non-typical ground motion characteristics.
\item Cascading rupture in fracture networks results in a significantly higher seismologically inferred stress drop.
\end{keypoints}

\section*{Introduction}
Geological faults develop complex multiscale features during their evolutionary stages, ranging from fine-scale topographic variations \cite{power1991euclidean, candela2010characterization} to large-scale branching and segmentation (i.e., fracture networks) \cite{chester1993internal, ben2003characterization, mitchell2009nature, mitchell2012towards}. These geometrical irregularities induce variations in stresses around the faults due to tectonic deformation and both aseismic and seismic events \cite{faulkner2010review, griffith2012coseismic}. They play a significant role in fault mechanics and earthquake rupture dynamics, including nucleation, propagation, and termination \cite{ide2005earthquakes, bhat2007role, dunham2011earthquake, okubo2019dynamics, kyriakopoulos2019dynamic}, resulting in distinct radiated seismic wavefields compared to simple fault geometries.

Although numerous studies have examined the role of geometrical complexity in dynamic rupture and its influence on the seismic wavefield, two fundamental questions remain unanswered: (1) How do cascading ruptures on complex fault geometries influence the generation of high-frequency seismic radiation? (2) How do cascading ruptures across multiscale fractures affect seismically inferred stress drop and near-source ground motion characteristics? Addressing these questions is essential for advancing our understanding of fracture-network cascading ruptures and for potentially improving seismic hazard assessment.

Complex fault geometries, such as fault bends, branches, or rough surfaces, create stress concentrations along the fault and may lead to high-frequency seismic wave radiation. When a rupture propagates through these areas, it undergoes rapid changes in stress conditions, resulting in abrupt acceleration or deceleration \cite{madariaga1977high, shi2013rupture}. Fault bends or step-overs may cause the rupture to change direction or split, resulting in fluctuations in rupture velocity and direction \cite{kame2008seismic}. Off-fault damage acts as an additional source of geometrical complexity and high-frequency radiation, occurring over small spatial and temporal scales \cite{andrews2005rupture, okubo2019dynamics, gabriel2021unified, zhao2024dynamic}.

Despite remarkable past efforts, observing and modeling the effects of complex fault geometry in detail remains a challenge. Common physics-based simulations often simplify the source as one or a few infinitesimally thin, planar surfaces. To approximate fault-zone complexity, modelers have used proxies such as lower rigidity zones \cite{huang2011pulse, huang2014earthquake}, fault roughness \cite{roten20113d, shi2013rupture, mai2018accounting}, and heterogeneous source properties \cite{ide2005earthquakes, ripperger2007earthquake, gallovivc2023broadband}. However, these proxies may not adequately capture the complete multiscale geometrical complexity observed in fault zones. The main difficulty lies in computational challenges and the lack of detailed fault models that explicitly incorporate multiscale fractures.

Over the past decade, facilitated by increased computation power and improved numerical methods, more advanced 3D dynamic rupture simulations that account for geometrical irregularity, earthquake source complexity, and heterogeneous media have demonstrated that these complexities can generate ground motions with high-frequency content \cite{withers2019validation, taufiqurrahman2022broadband, gallovivc2023broadband}. Physics-based simulations also suggest that the distribution of ground motions varies depending on earthquake rupture characteristics, highlighting the importance of conducting more realistic earthquake simulations, especially in the near-source region \cite{ide2005earthquakes, vyas2016distance, ando2017dynamic, wollherr2019landers}.

Current applications of these advancements include simulations that capture high-frequency ground motions up to 5–10 Hz in the near-source region by considering fault and material heterogeneity \cite{shi2013rupture, withers2019ground, taufiqurrahman2022broadband, vyas2024ground}. Recent studies by \citeA{palgunadi2024dynamic} and \citeA{gabriel2024fracture} show that using dynamic rupture simulations and models explicitly accounting for a geometrically complex fault network, moderate earthquakes may result from a cascading rupture across a fracture network, leading to distinct ground motion characteristics. Potential advantages of these models include a more accurate representation of rupture dynamics and improved calculation of ground-shaking intensity, which are critical for seismic hazard assessments.

Seismological estimates of corner frequency and stress drop are often based on the assumption of a relatively simple single-plane rupture model \cite{brune1970tectonic, madariaga1977high}. However, measurements, especially for moderate-sized earthquakes ($\Mw <$ 6), are not well resolved and contain significant random and potentially systematic uncertainties \cite{abercrombie2021resolution}. Unfortunately, real earthquake ruptures are more complex than the single-plane rupture model due to geometrical and source-related complexities. As indicated by their point-source representations of the source time function, earthquakes with complex source properties exhibit distinct displacement spectra, such as intermediate frequency fall-off \cite{abercrombie2021resolution}. Therefore, seismologically inferred corner frequency and stress drop estimates may only be applicable to earthquakes with relatively simple source geometries and properties. Investigating highly ideal cases using numerical modeling of fracture-network cascading earthquakes may provide insights into the consequence of the seismological approach to determine more complex earthquake source properties.

In this study, we leverage advanced 3D dynamic rupture simulations that explicitly incorporate geometrically complex, multiscale fractures to investigate how high-frequency ground motion is generated during cascading earthquakes. Unlike previous work, we model the fault network in detail, capturing realistic geometric features and using kinematic ground motion simulations to isolate the contributions of individual slipping fractures to the radiated seismic wavefield. We also analyze seismically inferred stress drops, investigate spatial variations in the frequency content of the near-source wavefield, and evaluate ground-shaking intensities. Our results are then validated against earthquake observations in Central Italy. By incorporating explicit modeling of multiscale fault complexity, this study offers new insights into the influence of fault geometry on ground motion characteristics.

The paper is organized as follows. In Section \colorednameref{section:modeling_setup}, we define our modeling strategy, describe the construction of the main fault and fracture network, and outline the dynamic rupture model ingredients. We then present the kinematic simulations that decompose the contribution of individual fractures to the overall seismogram. In Section \colorednameref{section:results}, we compare the ground motions from cascading ruptures to those from non-cascading scenarios and discuss the implications of these findings for seismic hazard assessment.

\section*{Modeling Setup}
\label{section:modeling_setup}
In this section, we summarize our approach to generating ground motions from dynamic ruptures occurring within a fracture network. For complete descriptions of the fracture network construction and the rupture parameters, we refer the reader to \citeA{palgunadi2024dynamic}. To examine the individual contributions of each fracture to the overall seismogram, we employ a kinematic simulation approach that utilizes rupture parameters derived from the outputs of dynamic rupture simulations.

\subsection*{Main Fault and Fracture Network Construction}

This study employs a fault model consisting of a listric fault with depth-dependent dip angles ranging from $30^\circ$ to $80^\circ$ and a strike direction of N270E relative to North. The main fault measures 8 km along-strike and 4 km along-dip, with a burial depth of 1 km, ensuring the fault does not intersect the free surface (Figure \ref{fig:model}). This type of listric fault geometry, common in transitional regimes between normal and strike-slip faulting, is often associated with subsurface reservoirs of oil, gas, or geothermal energy \cite{onajite2013seismic, ward2016reservoir, dixon2019geological}. The fractures are confined to a volumetric damage zone of 10 km $\times$ 1 km $\times$ 6 km in the $x$ (along-strike, East-West), $y$ (normal to the main fault, North-South), and $z$ (depth, positive upward) directions  (inset in Figure \ref{fig:slip}a).

The density and size distribution of the fractures are statistically constrained by field observations using power-law relationships \cite{mitchell2009nature, savage2011collateral}. Fracture density is quantified as the number of fractures per unit length ($P_{10}$), assumed to follow a power-law decay characterized by a decay constant $m=0.8$ and a coefficient $C=2.5$ as a function of fracture distance ($d$) ($P_{10}=2.5d^{-0.8}$ \cite{palgunadi2024dynamic}). Fractures are confined to the designated network region, with densities decreasing from the vicinity of the main fault outward. The fracture sizes ($R$) follow a power-law distribution with radii ranging from 100 m to 500 m, decaying as $\propto R^{-2}$ \cite{lavoine2019density}. The orientations of fractures are constrained by the observed orientations of large-scale fractures and conjugate faults within fault damage zones, typically ranging from $25^\circ$ to $30^\circ$ relative to the main fault \cite{mitchell2009nature}. The network includes two dominant fracture families with strike orientations of N120E and N20E, consisting of 423 and 431 fractures, respectively. Among all fractures, $78\%$ are connected to more than one fracture, $3\%$ are connected to only one fracture, and $19\%$ are unconnected (Figure \ref{fig:model} or Figure \ref{fig:slip}b).

\subsection*{Dynamic Rupture Simulations}

To conduct 3D dynamic rupture simulations, we use the open-source software SeisSol (see \colorednameref{section:data_resources}), which solves the dynamic rupture problem on a specified surface of potentially complex 3D geometry embedded in a 3D Earth-structural model and subjected to an initial regional stress field. These simulations allow modeling of the complete space-time dynamic earthquake rupture process in a physically self-consistent manner.

SeisSol employs fully adaptive, unstructured tetrahedral meshes, enabling the representation of geometrically complex geological structures \cite{pelties2014verification, uphoff2017extreme}. Recent computational optimizations version used in this study have significantly enhanced the accuracy and scalability of SeisSol \cite{uphoff2020yet, dorozhinskii2021seissol}, including through an efficient local time-stepping algorithm  \cite{breuer2016petascale, uphoff2017extreme}.

The computational domain for this study is a $40 \times 40 \times 20$ km$^3$ cube, discretized into an unstructured tetrahedral mesh with varying element edge lengths. We employ high-order basis functions of polynomial degree $p = 4$, achieving fifth-order accuracy ($\mathcal{O}5$) in both space and time for seismic wave propagation. To ensure the resolution of the minimum breakdown zone size, the smallest fracture (100~m) is resolved with a maximum on-fault resolution of 4~m, resulting in a minimum cohesive zone width of 8~m (following \citeA{wollherr2018off}). This is achieved using an average of two fifth-order accurate elements, each with 12 Gaussian integration points. The mesh is refined within a smaller cube measuring $26 \times 26 \times 10$ km$^3$, with the largest mesh element edge length set at 250 m, resulting in over 88 million tetrahedral cells. We model the medium as a homogeneous 3D half-space with P-wave velocity $v_p = 6.0$~km/s, S-wave velocity $c_s = 3.464$~km/s, rock density $\rho = 2670$~kg/m$^3$, and rigidity $\mu = \lambda = 32$~GPa. This mesh configuration resolves frequencies up to 7 Hz everywhere within the refined volume and approximately 13 Hz at all stations close to the fracture network.  

We adopt a laboratory-based rate-and-state friction law \cite{rice2006heating} with rapid velocity weakening \cite{noda2009, dunham2011earthquake}. The frictional parameters are held constant, except for the state evolution slip distance $L$, which scales linearly with fracture size \cite{gabriel2024fracture}. The model includes the direct effect parameter ($a=0.01$) and the evolution parameter ($b=0.014$), ensuring the main fault and fractures remain frictionally unstable. The characteristic weakening velocity $V_w = 0.1$~m/s is based on the experimentally observed set of rapid decay of the effective friction coefficient, ranging between 0.01 and 1 m/s \cite{rice2006heating, beeler2008constitutive, di2011fault}. The steady-state friction coefficient ($f_0$) is fixed at 0.6 at a reference slip velocity $V_0=10^{-6}$~m/s, with a weakened steady-state friction coefficient ($f_w$) of 0.1 \cite{rice2006heating}. 

In this study, we focus on three specific cases described in \citeA{palgunadi2024dynamic}: \textit{Case 1}, a pure cascading rupture within the fracture network, is obtained by assuming a maximum horizontal stress orientation ($\Psi$) of $65^\circ$. \textit{Case 2}, a rupture with off-fault fracture slip and \textit{Case 3}, a main-fault rupture without the fracture network in the model, are obtained with $\Psi=120^\circ$, as detailed in \citeA{palgunadi2024dynamic}. Fault and fracture strength are characterized using the relative pre-stress ratio $\mathcal{R}$ \cite{aochi20031999,ulrich2019dynamic}.  
$\mathcal{R}$ is defined as $(\tau_0 - f_w \sigma'_n)/((f_p - f_w)\sigma'_n)$, where $\tau_0$ is the initial shear stress, $\sigma'_n$ is the effective normal stress, and $f_p$ denotes the peak friction coefficient during dynamic rupture, which here is approximated by its reference value $f_0$. A constant $\mathcal{R}_0 = 0.8$ is set for fractures at the maximum possible value of $\mathcal{R}$, with shear stress to effective normal stress loading ratios ($\tau_0/\sigma_n'$) ranging from 0.2 to 0.5, indicating sub-critical stress conditions.

\textit{Case 1} features a favorable relative pre-stress ratio ($\mathcal{R} \geq 0.6$) for the fractures and unfavorable relative pre-stress ratio ($\mathcal{R} < 0.3$) for the main fault ($\Psi=65^\circ$); \textit{Case 2} favorable $\mathcal{R}$ for the main fault but unfavorable $\mathcal{R}$ for fractures ($\Psi=120^\circ$); \textit{Case 3} has the same stress conditions as \textit{Case 2} but does not incorporate the fracture network, hence, the rupture is restricted to the single main fault surface. Here, we focus our attention on the radiated seismic wavefield and the implications of the intricate rupture dynamics on the ground motion properties.

\subsection*{Kinematic Ground Motion Simulation}
\label{subsection:kinematic_rupture}
Throughout this study, we use the synthetic waveforms from the dynamic rupture models and analyze characteristic frequencies and other ground motion parameters. Furthermore, we perform kinematic ground motion calculations to explore the contribution of individual fractures to the overall radiated seismic wavefield.

To understand which slipped fracture contributed to which parts of the radiated seismic wavefield, it is necessary to decompose the complete wavefield. However, in dynamic rupture simulations, decomposition of the seismic wavefield into the individual contributions of the slipped fractures is impossible due to the inherent non-linearity. Therefore, we employ kinematic rupture simulations as a means to efficiently decompose the dynamic rupture waveforms into individual contributions (Figure \ref{fig:Schematic}).

For the kinematic ground motion computation, we use the open-source software AXITRA \cite{cotton1997dynamic}, which models wave propagation from the source to the receiver generated by a prescribed earthquake rupture model. AXITRA computes the stress field radiated by the six components of the moment tensor, employing the reflectivity method alongside discrete wavenumber decomposition Green’s functions for an axis-symmetric medium \cite{bouchon1981simple}. In our study, we apply the kinematic source parameters as obtained from the dynamic rupture simulations, including slip (see Figure \ref{fig:slip}), moment-rate function, seismic moment, rake angle, and rise time (refer to Figure \ref{fig:CompareSeismogram} and \ref{fig:riseTime}). 

The Green’s functions assume a homogeneous medium, consistent with the dynamic rupture model, and target a frequency of 13 Hz. Each fault element (triangle), with a minimum slip of 0.001 m, is treated as a point source with its individual kinematic parameters and source time functions. The seismogram generated from the kinematic model integrates waveform contributions from all slipped fractures and the main fault. For the fracture-network cascading rupture scenario, the kinematic source model incorporates approximately 8 million point sources, corresponding to each slipped cell on the fractures and the main fault. For other scenarios, the kinematic rupture is sampled with 5 million and 1.5 million point sources, respectively, aligning with cases involving rupture with (\textit{Case 2}) and without off-fault fracture slip (\textit{Case 3}).

\subsection*{Equivalent Near-Field Characteristic Frequency}\label{subsection:corner_frequency}
We measure the equivalent near-field characteristic frequency ($f_c$, \citeA{schliwa2024equivalent}) to analyze specific source characteristics in the resulting seismic waveforms. The physical size and characteristics of earthquake sources are not directly observable and are typically inferred from seismic waveforms in the near- and far-field and/or ground-displacement data using an earthquake source-inversion approach \cite{ji2003slip, melgar2015kinematic, mai2016earthquake, vasyura2020bayesian}. In addition, far-field seismic waveforms and their Fourier amplitude displacement spectra are used to infer point-source rupture parameters. This inference involves estimating $f_c$ by minimizing the misfit between the observed displacement spectra and Brune’s model, e.g., \citeA{kaneko2014seismic} as: 

\begin{equation}\label{eq:fc}
    u(f) = \frac{\Omega_0}{\left(1 + (f/f_c)^{\gamma n} \right)^{1/\gamma}}\,,
\end{equation}
where $u(f)$ represents the waveform displacement spectrum, $\Omega_0$ is the long-period spectral level related to the total seismic moment of the earthquake, $\gamma$ defines the corner shape, and $n$ denotes the spectral fall-off. 

Following \citeA{schliwa2024equivalent}, we use Brune’s model to examine the spectral-decay characteristic of the near-field waveforms, recognizing that Eq. \ref{eq:fc} has been developed under the far-field assumption. However, as proposed in \citeA{schliwa2024equivalent}, the spatial variations of near-field amplitude spectra may carry important information on the details of the rupture. We calculate $f_c$ for each virtual receiver at the Earth's surface to assess potential spatial variations in $f_c$.
For simplicity, we set the high-frequency fall-off $n=2$ and fix the corner shape to solve Eq. \ref{eq:fc} \cite{brune1970tectonic}. It is important to note that in the idealized setup of this study, neither intrinsic nor scattering attenuation are modeled, nor are site effects considered. Consequently, the spectral shape of the simulated ground motions reflects only source effects and geometrical spreading in a homogeneous half-space.
We explored both the original Brune model ($\gamma=1$) and the sharper \citeA{boatwright1980spectral} model ($\gamma=2$), finding that the former provides a better fit. We focus on station distances less than 15 km and determine $f_c$ using the grid search method of \citeA{schliwa2024equivalent}, the best fit's $95\%$ confidence interval considered as the uncertainty in $f_c$.

\section*{Results}
\label{section:results}
The cascading earthquake generates pulses on the main fault and fractures in the macroscale view, distinguished by their short rise time compared to the rupture time (see Figures \ref{fig:riseTime}a and \ref{fig:riseTime}b) \cite{heaton1990evidence} and slipped fractures across the fracture network (see Figure \ref{fig:slip}). The cascading rupture generates localized high slip patches at several fault-fracture or fracture-fracture intersections, as detailed in \citeA{palgunadi2024dynamic}. In contrast, the non-cascading earthquake produces slip concentrated on the main fault and its directly connected fractures.  

To analyze differences in seismograms or ground motions, we compare three scenarios: (1) a cascading rupture within the fracture network (\textit{Case 1}), (2) a rupture with off-fault fracture slip (\textit{Case 2}), and (3) a main-fault rupture without the fracture network being incorporated into the model (\textit{Case 3}), resulting in magnitudes of $\Mw$ 5.5, $\Mw$ 6.0, and $\Mw$ 6.0, respectively. Figure \ref{fig:WaveformSpectral} displays three-component velocity seismograms for all three cases at five stations, respectively. \textit{Cases 1} and \textit{2} are discussed in detail in \citeA{palgunadi2024dynamic}.

The following analysis of the radiated seismic wavefield and its ground-motion properties is divided into three parts. In section \colorednameref{subsection:near_field_seismogram}, we explore the characteristics of near-field seismograms and the contributions of each fracture to the seismogram. In this section, we use the kinematic approach to decompose the contribution of each fracture. Section \colorednameref{subsection:azimuthal_dependence_fc} investigates the distribution of equivalent near-field characteristic frequencies for both cascading and non-cascading earthquake scenarios. This section uses the dynamic rupture model waveform outputs. Lastly, Section  \colorednameref{subsection:ground_motion_characteristics} presents results on strong ground motion parameters such as peak ground acceleration (PGA), peak ground velocity (PGV), and pseudo-spectral acceleration at selected periods $T$ and based on $5\%$ damping (PSA($T$)). We examine synthetic seismograms from dynamic rupture model outputs at 81 near-field stations located within a 12.5 km radius from the center-top of the main fault.

\subsection*{Near-Field Seismic Wavefields from Dynamic and Kinematic Rupture Models}
\label{subsection:near_field_seismogram}

\subsubsection*{Comparative Analysis of Ground-Motion Properties}
To compare the three cases, we use the outputs generated by dynamic rupture models. Figure \ref{fig:WaveformSpectral}a highlights significant differences in ground-motion properties between fracture-network cascading (\textit{Case 1}) and non-cascading earthquakes (\textit{Cases 2} and \textit{3}), as described in \citeA{palgunadi2024dynamic}. Notably, \textit{Case 1} exhibits higher frequency content compared to \textit{Cases 2} and \textit{3}, as illustrated in Figure \ref{fig:WaveformSpectral}b. The waveforms of \textit{Case 1} are characterized by short-wavelength amplitude variations across all components and stations (highlighted in red in Figure \ref{fig:WaveformSpectral}a). Specifically, the East-West (fault-parallel) component shows higher amplitudes than the North-South (fault-normal) component. Additionally, the waveforms in \textit{Case 1} include late-arriving waves that resemble ``coda” waves typically generated by seismic scattering, here attributed solely to rupture (i.e., source) complexity and associated increased rupture duration compared to \textit{Cases 2} and \textit{3} (Figure \ref{fig:Schematic}).

The seismograms for \textit{Cases 2} and \textit{3} display nearly identical low-frequency signatures across all components. However, \textit{Case 3} has lower frequency content than \textit{Case 2}, attributed to the dynamic rupture that occurs only on the main fault, without involving the fracture network in the model (depicted in blue in Figure \ref{fig:WaveformSpectral}b). In both cases, higher amplitudes are observed on the vertical component (UD) at stations 2 and 81, located on the hanging wall of the listric main fault (Figure \ref{fig:WaveformSpectral}a). 

\subsubsection*{Validation and Detailed Analysis of High-Frequency Content}
To investigate the origin of the source coda wave-like signatures observed in \textit{Case 1}, we reconstruct the overall seismograms by combining contributions from each fracture. Dynamic rupture simulations are non-linear, which does not permit the direct isolation of contributions from individual seismic sources (i.e., ruptured fractures) within a fracture network. While isochrone analysis is possible for simpler fault geometry \cite{spudich1984direct,schliwa2024equivalent}, we here conduct kinematic simulations to isolate individual fracture contributions from the seismograms. This method is commonly used to generate the seismic wavefield from dynamic rupture simulations (e.g., from dynamic rupture models based on the boundary integral element method \cite{aochi2002three, goto2010simulation} or for generating teleseismic synthetics \cite{vanDriel2015, Taufiq2023}) and was applied by \citeA{ripperger2007earthquake, mai2018accounting, gallovivc2023broadband}. All elements on the slipped fractures and the main fault are treated as point sources with their respective source parameters.

To validate the kinematic approach, we present a comparison of seismograms generated directly in a dynamic rupture model (shown in solid red line) and through kinematic rupture (shown in dashed blue line) for \textit{Case 1}, specifically for near-field stations across the North-South direction (red circle in the inset of Figure \ref{fig:CompareSeismogram}). The kinematic approach well reproduces the wave characteristics observed in the dynamic rupture seismograms in the frequency range 0.001 - 6 Hz. Figure \ref{fig:Schematic} illustrates a sketch of how different parts of the rupture radiate seismic waves, and how these then combine to form the complete seismograms at a site of interest (Figure \ref{fig:CompareSeismogram}). 

The first wave arrivals correspond to the P-wave, marking the onset of the fracture-network cascading rupture. The subsequent S-wave arrivals are not clearly distinguishable. The distinctive source coda wave-like signatures arise from sequences of fractures that slip later in the shaking sequence, leading to a gradual decrease in amplitude as the event progresses.

\subsection*{Azimuthal-Dependence of Equivalent Near-Field Characteristic Frequency}
\label{subsection:azimuthal_dependence_fc}
The equivalent near-field characteristic frequency ($f_c$, \cite{schliwa2024equivalent}) is calculated using Brune's model, as described above in the section \colorednameref{subsection:corner_frequency}. This frequency is crucial for characterizing earthquake sources by analyzing seismic waveform spectra \cite{kaneko2014seismic, trugman2017application}. We assess and compare the waveform spectra and $f_c$ across the cascading and non-cascading ruptures, providing statistics and spatial distributions of $f_c$.

\subsubsection*{Frequency Content}
Figure \ref{fig:FcAzimuth} shows $f_c$ values for the fracture-network cascading earthquake (\textit{Case 1}) for the vertical component; EW and NS components are depicted in Figures \ref{fig:FcDistributionEW} and \ref{fig:FcDistributionNS}, respectively. On average, \textit{Case 1} generates a broad $f_c$ distribution with a mean of 1.1 Hz and a standard deviation of 0.5 Hz (Figure \ref{fig:fc_statistics}a). Comparable mean and standard deviation values are observed across all components. In contrast, \textit{Cases 2} and \textit{3} exhibit a narrower distribution of $f_c$ values, predominantly below 0.5 Hz. \textit{Case 2} shows a higher average $f_c$ than Case 3, with standard deviations of 0.09 Hz (Figure \ref{fig:fc_statistics}b) and 0.07 Hz (Figure \ref{fig:fc_statistics}c), respectively. 

Despite these variations, the mean $f_c$ values from all three cases align with estimates based on \citeA{allmann2009global} (Figure \ref{fig:FcAzimuth}b). The seismologically inferred stress drop estimates for \textit{Cases 2} and \textit{3} closely match the mean values from the simulations. For the cascading rupture \textit{Case 1}, the mean $f_c$ is higher than the typical value for a moment magnitude of $\Mw$ 5.5. It corresponds to an estimated stress drop of approximately 75 MPa—which is almost an order of magnitude higher than the average stress drop directly calculated on the fractures and fault, which is 9.6 MPa \cite{palgunadi2024dynamic}. This result suggests that in the near-field region, spectral characteristics are sensitive to multi-fracture rupture behavior. Hence, the average stress drop calculation estimated by Brune's model across all stations is biased by slip on the fractures, and associated high $f_c$ observed at stations with azimuths aligned with the fracture trends. Although the average stress drop calculated on the fractures and main fault is 9.6 MPa, on a local scale, some patches on a fracture experience a maximum stress drop of up to 12 MPa, especially at the intersections between two or more fractures \cite{palgunadi2024dynamic}.

\subsubsection*{Azimuthal Variations and Validation of $f_c$}

We extract $f_c$ values within a 4.8 and 5.2 km distance from the epicenter for all azimuths to examine their azimuthal dependence. The highest $f_c$ values, reaching up to 5 Hz, are observed in the vertical component near the epicenter. $f_c$ demonstrates an azimuthal dependence \cite{KanekoShearer2015}, with higher frequencies in the direction of the slipped fracture network (Figure \ref{fig:FcAzimuth}c). Some $f_c$ estimations, particularly those exceeding 4 Hz, show high uncertainties ($\Delta f_c \sim$ 2 Hz). On average, azimuths of around $20^\circ$, $120^\circ$, $220^\circ$, and $300^\circ$ correspond to high $f_c$ values ($f_c \sim$ 4 Hz). These azimuths approximately align with the fault planes of the equivalent point-source moment tensor solution, characterized by pure strike-slip faulting mechanisms (strike = $114^\circ$, dip = $88^\circ$, rake = $0^\circ$) in \textit{Case 1} (Table 2 in \citeA{palgunadi2024dynamic}, with $\Psi = 65^\circ$).

We put into perspective the obtained distribution of corner frequency by analyzing real earthquake recordings from the three largest earthquakes of the well-documented recent Central Apennines earthquake sequence, in Italy: the 2009 $\Mw$ 6.2 L'Aquila earthquake, the 2016 $\Mw$ 6.2 Central Italy (Amatrice) earthquake, and the 2016 $\Mw$ 6.6 Norcia earthquake. These events were selected due to the dense distribution of strong-motion seismometers and the availability of freely accessible data (see \colorednameref{section:data_resources}). To calculate corner frequency, each seismogram is analyzed using Brune’s model. All three components are treated separately and then averaged. We identify an azimuthal dependence of the equivalent near-field characteristic frequency for all three earthquakes: the Norcia earthquake (see Figure \ref{fig:fc_norcia}), the Amatrice earthquake (Figure \ref{fig:fc_realEarthquake}a), and the L’Aquila earthquake (Figure \ref{fig:fc_realEarthquake}b). Despite variations in the epicentral distances of the stations, higher corner frequency values tend to be oriented toward the fault strike. For these earthquakes, the corner frequency ranges from 0.1 to 0.85 Hz, with an average of approximately 0.3 to 0.4 Hz \cite{maercklin2011effectiveness, calderoni2017rupture, supino2019probabilistic}. 

Based on the synthetic data from \textit{Case 1} and observed strong motion data, the radiation pattern and forward-directivity may lead to an increase in the corner frequency or $f_c$ in the fracture direction, at least for stations within an epicentral distance of less than 100 km. This observation is particularly emphasized for stations close to the earthquake source and in agreement with large-scale dynamic rupture simulations results in \citeA{schliwa2024equivalent}. 

\subsubsection*{Map of Azimuthal Variations of $f_c$}
We analyze the relationship between the equivalent near-field characteristic frequency ($f_c$) and source complexity. Figure \ref{fig:CornerFreq} illustrates that $f_c$ values vary across different azimuths, with higher frequencies observed in the dominant fracture strike directions. The higher frequencies inferred from the vertical component correlate with the SV-wave radiation pattern (Figure \ref{fig:CornerFreq}a), representing the nodal lobes where seismic energy consists of incoherent radiation scattered by the randomized fracture orientations. The relatively high $f_c$ persists up to a resolved distance of 12.5 km across all azimuths. Conversely, the distribution of $f_c$ values in the East-West and North-South components appears less pronounced. 

High-resolution dynamic rupture simulations for the 1992 $\Mw$ 7.3 Landers earthquake and the 2019 $\Mw$ 7.1 Ridgecrest earthquake also reveal spatial variations in corner frequency or $f_c$ \cite{schliwa2024equivalent}. $f_c$ variations are caused by the radiation pattern, rupture directivity, and local dynamic source effects, e.g., facilitating the identification of rupture gaps and rake-rotations due to free-surface interaction (e.g., \citeA{kearse2019curved}). Our findings further emphasize the consistency across various seismic events for using near-field $f_c$ variability for source characterization. 

The presence of two pairs of peaks corresponding to nearly orthogonal fracture families indicates the characteristics of the cascading rupture across these two dominant fracture families. This $f_c$ pattern is also reflected in the moment tensor solution of \textit{Case 1}, which is misaligned with the strike of the main fault \cite{palgunadi2024dynamic, gabriel2024fracture}.

\subsection*{Ground Motion Characteristics}
\label{subsection:ground_motion_characteristics}

\subsubsection*{Comparing simulated ground-motions to empirical ground-motion models}

We apply a site-amplification correction to the ground motion synthetics before comparing our simulations with Ground Motion Models (GMMs). This correction accounts for the fact that our simulations are conducted in a homogeneous high shear wave velocity ($c_s$ of 3.464 km/s), in contrast to data-based GMMs that include site conditions that most often are characterized by much lower seismic wave speeds ($V_{S30} < 1.5$ km/s). We apply an amplification factor to all seismograms to scale amplitudes to represent an assumed $V_{S30}$ value of  0.76 km/s, following the methodology outlined by \citeA{borcherdt1994estimates, borcherdt2002empirical}. This approach has been applied in previous studies (e.g., \citeA{mai2010hybrid, moczo2018key}). Figure \ref{fig:corrections} illustrates the filter and its effect on ground motion waveforms. The filter amplifies ground motion in the frequency range of 0.1 Hz to 3 Hz, while slightly decreasing amplitudes above 3 Hz (Figures \ref{fig:corrections}a and \ref{fig:corrections}b). The waveform differences before and after correction are minimal (Figure \ref{fig:corrections}c).

We then compare the PGV and PGA of 81 stations against the predictions of selected GMMs (Figure \ref{fig:GMM}) by \citeA{abrahamson2014summary}, \citeA{boore2014nga}, \citeA{campbell2014nga}, and \citeA{chiou2014update}. The three cases exhibit distinct peak amplitude characteristics. In \textit{Case 1}, the PGA values are generally higher than the median GMM predictions but remain within the GMMs' standard deviation. Some values, particularly at stations very close to the main fault ($R_{JB} < 5$~km), tend to overestimate the GMMs. In \textit{Case 2}, the PGA values are slightly below the GMMs' median value for an $\Mw 6.0$ (represented by dashed lines in Figure \ref{fig:GMM}a). In \textit{Case 3}, the PGA values generally fall below the median GMM predictions, indicating an overall underestimation of the synthetics.

For PGV, \textit{Case 1} aligns well with the median GMM value for $R_{JB} > 5$~km and underestimates GMMs for $R_{JB} \leq 5$~km. \textit{Cases 2} and \textit{3} yield comparable PGV values compared to GMMs. Both cases show two clusters of PGV values, corresponding to ground shaking on the hanging wall and footwall, respectively. The lower PGV values may be attributed to the fact that our simulations consider a homogeneous velocity model and small-scale source effects, while GMMs incorporate data recorded under a wide range of geological conditions. Figure \ref{fig:GMM} confirms that the ground motions from our 3D dynamic rupture simulations are in general agreement with empirical predictions. We, therefore, proceed to discuss directional patterns of shaking and further ground-motion properties.

We assess PGA and PGV values in different azimuthal directions and compare these with GMMs to examine azimuthal dependencies of shaking levels (Figure \ref{fig:pga_pgv_azimuth}). The azimuth angles range from $0^\circ$ to $360^\circ$, increasing clockwise with $0^\circ$ corresponding to North. In \textit{Case 1}, a clear distinction is observed between the violet and green dots, representing azimuthal angles of approximately $300^\circ$ and $120^\circ$, respectively. The PGA and PGV values in the rupture direction ($120^\circ$, green squares) tend to overestimate the GMMs for $R_{JB} > 5$~km, while the region corresponding to the backward rupture direction aligns well with the GMMs (violet squares) and is underestimated for $R_{JB} < 5$~km. In \textit{Cases 2} and \textit{3}, the PGA values generally underestimate the GMMs, except in specific azimuthal directions ($110^\circ - 130^\circ$) where simulated ground motions exceed GMM predictions at distances beyond 10 km. The PGV values show good agreement with the GMMs in the hanging wall region.

\subsubsection*{PGV and PGA Shake Maps}

Ground motions exhibit complexity due to variations in the spatio-temporal evolution of ruptures, as demonstrated in various studies (e.g., \citeA{mai2009ground, dunham2011earthquake, shi2013rupture}). The cascading rupture, involving complicated dynamic interactions within the fracture network \cite{palgunadi2024dynamic}, results in ground motions that significantly differ from those of non-cascading ruptures. Figure \ref{fig:PGA} presents maps comparing PGV using GMRotD50, a metric independent of the sensor orientations \cite{boore2006orientation} for three scenarios: cascading rupture (\textit{Case 1}, Figures \ref{fig:PGA}a and \ref{fig:PGA}d), rupture with off-fault fractures slip (\textit{Case 2}, Figures \ref{fig:PGA}b and \ref{fig:PGA}e), and rupture without the fracture network in the model (\textit{Case 3}, Figures \ref{fig:PGA}c and \ref{fig:PGA}f). 

\textit{Case 1} exhibits greater variability in PGA and PGV amplitudes across all azimuths compared to \textit{Cases 2} and \textit{3}. The distributions for \textit{Cases 2} and \textit{3} indicate pronounced shaking on the hanging wall side of the listric fault, corroborating findings by \citeA{abrahamson1993estimation} and \citeA{somerville1996implications}. This observation is consistent with simulations of ground shaking along listric faults \cite{ofoegbu1998mechanical, passone2017kinematic, rodgers2019effect, moratto2023near}. 

The variability in PGA for \textit{Case 2} is largely attributable to off-fault fracture slip, evident from the relatively higher PGA values on the footwall side compared to \textit{Case 3} (Figures \ref{fig:PGA}b and \ref{fig:PGA}c). While both \textit{Cases 1} and \textit{2} involve multiple fractures, the amplitude variability of PGA between the two cases is significant. \textit{Case 1}, with 561 slipped fractures representing $70\%$ of the total fractures, exhibits higher PGA values. In contrast, \textit{Case 2}, with 477 slipped fractures accounting for $58\%$ of the total, shows less variability. This less than $20\%$ increase in the number of slipped multi-scale fractures results in a substantial rise in PGA values. This dramatic difference in PGA is mainly due to a significant increase in rupture speed ($V_r$) within individual fractures (Figure \ref{fig:rupture_speed_statistics}). In \textit{Case 1}, the rupture speed is predominantly $V_r > 2900$~m/s (Figure \ref{fig:rupture_speed_statistics}a), whereas \textit{Case 2} features slower $V_r$ values (Figure \ref{fig:rupture_speed_statistics}b). According to the kinematic approach, $V_r$ plays a major role in producing high peak amplitudes, especially in the near-field region \cite{robinson2006mw, zollo2006earthquake, bizzarri2008effects}.

\subsubsection*{Azimuthal dependence of ground motions}
Our comparative study of \textit{Cases 1}, \textit{2}, and \textit{3} reveals significant differences in the azimuthal dependence of ground motion intensities for the near-field region, highlighting ground motion patterns between cascading and non-cascading ruptures. To further analyze the distribution of PGA and PGV values across all azimuths, we calculate the mean and standard deviation for every $10^\circ$ azimuthal bin for an epicentral distance of $< 12.5$~km. In Figures \ref{fig:azimuthal_groundMotions}a and \ref{fig:azimuthal_groundMotions}b, the red color represents \textit{Case 1}, which shows no apparent azimuthal dependence of PGA and PGV values. Instead, they remain relatively constant across all azimuths. \textit{Case 1} suggests that ground motions are uniformly distributed across all azimuths. In contrast, \textit{Case 2} and \textit{Case 3} exhibit clear azimuthal dependence and demonstrate similar patterns. The green and blue colors in Figures \ref{fig:azimuthal_groundMotions}a and \ref{fig:azimuthal_groundMotions}b indicate high PGA and PGV values on the hanging wall ($270^\circ<$ azimuth $\leq 360^\circ$ and $0^\circ\leq$ azimuth $< 90^\circ$) and low values on the footwall ($90^\circ \leq$ azimuth $\leq 270^\circ$). Interestingly, the standard deviation of \textit{Case 1} is significantly higher on the footwall side compared to \textit{Case 2} and \textit{Case 3} for both PGA and PGV.

\subsubsection*{Pseudo-spectral acceleration}
Pseudo-spectral acceleration (PSA) is a widely used measure for assessing the severity of ground motion on structural systems in earthquake engineering. Based on single-degree-of-freedom spring-mass-damping systems, it significantly aids in estimating the peak response of structures to seismic events \cite{baker2021seismic}. We calculate PSA values with a damping factor of $\zeta = 5\%$ over a period range of $T = 0.1 - 10$~seconds. We then compare these PSA values from the three scenarios against several ground motion prediction models (GMMs) \cite{ abrahamson1997empirical, boore1997equations, boore2014nga, campbell2014nga}.

Figure \ref{fig:SAgraph}a compares our obtained PSA values against the GMMs. \textit{Case 1} PSA values overall align with the GMMs' median values, but exhibits higher values at shorter periods and lower values at longer periods, compared to the GMMs. Specifically, \textit{Case 1} shows elevated PSA values for shorter periods ($T=0.1 - 0.2$ s), indicative of intense high-frequency ground-motion radiation. In contrast, \textit{Cases 2} and \textit{3} demonstrate higher PSA values at longer periods and lower PSA values at shorter periods relative to the median GMM values. This pattern is highlighted in Figure \ref{fig:SAgraph}a, with \textit{Case 2} (green dots) showing higher PSA values at short periods compared to \textit{Case 3} (blue squares). However, at longer periods ($T > 1$ s), PSA values for \textit{Cases 2} and \textit{3} converge, suggesting that seismic wave radiation from the main fault predominates at these periods, while contributions from slipping off-fault fractures are more significant at shorter periods.

Figures \ref{fig:SAgraph}b, \ref{fig:SAgraph}c, and \ref{fig:SAgraph}d illustrate the distance-dependent characteristics of PSA for each case over various periods. PSA values for the three cases align well with respective GMMs ($\Mw$ 5.5 for the fracture-network cascading earthquake and $\Mw$ 6.0 for \textit{Case 2} and \textit{Case 3}). At longer periods, the PSA values for all cases are comparable and fall within the median estimates provided by the GMMs.

Figure \ref{fig:SAmap} explores the distribution of PSA values at $T=0.3$ s, shedding light on the extent of ground shaking across different azimuths. In \textit{Case 1}, significant ground shaking is evident across all azimuths, much like the observations for PGA and PGV in Figure \ref{fig:PGA}. PSA values at $T=0.3$ s exhibit relatively constant levels across all directions, as depicted by the red dots in Figure \ref{fig:azimuthal_groundMotions}c, with consistent standard deviations.

For \textit{Cases 2} and \textit{3}, intense ground shaking primarily occurs on the hanging wall side of the main fault. Notably, \textit{Case 2} also displays elevated PSA values to the southeast of the epicenter, reflecting the activation of off-fault fractures within the damage zone. This spatial distribution underscores the impact of off-fault fracture activation in generating high ground-motion amplitudes across various azimuths. Both \textit{Cases 2} and \textit{3} show higher PSA values on the hanging wall side and reduced values towards the footwall. Intriguingly, \textit{Case 2} exhibits higher PSA values on the footwall side than \textit{Case 3}, presumably owing to the additional off-fault slipping fractures (represented by green triangles in Figure \ref{fig:azimuthal_groundMotions}c).

\section*{Discussion}
\label{section:discussions}
Our high-resolution 3D dynamic rupture cascade simulations within a geometrically complex fracture network provide valuable insights into the characteristics of high-frequency seismic wave generation and radiation. In this section, we discuss the properties of seismic waves resulting from fracture-network cascading ruptures, exploring their implications, significance, and inherent limitations.

\subsection*{Seismogram Characteristics of a Fracture-Network Cascading Rupture}\label{section:observation}

Cascading earthquakes within a fracture network generate high-frequency seismograms across all components and distances, as demonstrated in our near-field numerical simulations. However, for real earthquakes, this high-frequency content is further influenced by complex Earth structures such as medium heterogeneity, fault roughness, and topography, echoing findings from previous studies \cite{bouckovalas2005numerical, sato2012seismic, imperatori2015role, vyas2021characterizing, park2022correction, taufiqurrahman2022broadband, gallovivc2023broadband}. In practical observations, discerning whether an earthquake has occurred as a rupture cascade remains challenging due to Earth’s heterogeneities that obscure the origins of high-frequency radiated energy.

The cascading rupture is marked by an indistinct onset of S-waves across all components, attributed to the concurrent arrival of multiple P- and S-waves generated by slip on the various fractures, as demonstrated by our kinematic approach. As these waves propagate over longer distances, the distinction between wave types becomes clearer owing to increased travel times (station 81 versus 45 in Figure \ref{fig:WaveformSpectral}).

\subsection*{High $f_c$ and Stress Drop for a Fracture-Network Cascading Rupture}
Our numerical simulations (both dynamic and kinematic approaches) indicate that fracture-network cascading earthquakes generate seismic waves with higher equivalent near-field characteristic frequencies ($f_c$). The distribution of $f_c$ values across different azimuths serves as an indicator for identifying the characteristic behavior of fracture-network cascading earthquakes. Our findings show that $f_c$ values demonstrate an azimuthal dependence, aligning with the radiation pattern of the fracture direction \cite{trugman2021earthquake, schliwa2024equivalent}. While possible as in this study, confirming this azimuthal dependence in actual observations which have limited recording networks poses practical challenges due to 3D scattering in heterogeneous media, topographic effects, site-specific conditions, and intrinsic attenuation.

To further validate our findings on the azimuthal variations of $f_c$ as a proxy for corner frequency, employing near-field measurements from a station array of diverse azimuths would be beneficial, such as those available from dense seismic networks like Italy’s strong-motion network (see Figure \ref{fig:fc_realEarthquake}). Moreover, as seismic instrumentation becomes increasingly denser and more widely available, observations of spatially varying corner frequencies could provide critical insights into rupture propagation and directivity effects, thereby enhancing our understanding of earthquake source mechanics.

The fracture-network cascading rupture is also characterized by a higher average $f_c$. The far-field corner frequency estimated using seismologically inferred methods (Brune's model spectra) is considerably higher than expected based on moment magnitude scaling relations (Figure \ref{fig:FcAzimuth}c). Consequently, this higher average $f_c$ corresponds to a high stress drop, which is estimated to differ by almost an order of magnitude compared to the average stress drop calculated directly on the fault from the dynamic rupture simulation. In real earthquake events, direct on-fault estimation of stress drop is not possible. Nevertheless, our simulations suggest that using the conventional single-plane rupture model is not appropriate for complex earthquake sources like rupture cascades. In the conventional model (i.e., seismologically inferred $f_c$), the stress drop is proportional to $M/R^3$, where the source radius $R \sim c_s/f_c$, with $M$ being the seismic moment and $c_s$ the S-wave speed. In this relation, the estimation of $M$ is presumably valid, while the inferred source radius is more representative of a single fracture or fault plane, not the collective rupture of a fracture network cascade involving hundreds of fractures. Nevertheless, this high average $f_c$ and stress drop using Brune's model estimation could be potentially observable features of fracture-network cascading ruptures.

\subsection*{Implication of a Fracture-Network Cascading Rupture to Seismic Hazards}

The ground shaking generated by fracture-network cascading and non-cascading ruptures exhibits distinct spatial amplitude patterns. Cascading ruptures result in heterogeneously distributed strong ground shaking across all azimuths, whereas non-cascading ruptures typically generate intense shaking concentrated on the hanging wall side of the listric fault. Because the orientation of maximum ambient horizontal stress ($\Psi$) plays a critical role in dictating the rupture process, it also determines the actual shaking values and their distributions given a specific fault and fault system under consideration. In essence, for improved ground motion estimation, the regional stress field must be considered. However, accurately determining local stress orientations remains challenging due to significant uncertainties in current estimates \cite{heidbach2018world}. As indicated by \citeA{palgunadi2024dynamic}, a slight $20^\circ$ difference in $\Psi$ can shift the rupture style from cascading to non-cascading within the fracture network.

Fracture-network cascading ruptures radiate more high-frequency seismic energy compared to ruptures with off-fault fracture slip, suggesting that current ground motion models might underestimate PGA-values, especially near faults surrounded by a multi-scale fracture network. Our simulations indicate that the average PGA in cascading cases is nearly twice as large as that of non-cascading cases, highlighting the limited effectiveness of non-cascading ruptures in radiating high-frequency waves. Additionally, the activation of fractures limits the extent and affected area of high PGV values. Therefore, incorporating measurable geometrical fault complexity may improve the accuracy of ground-shaking estimates. 

The presence of strong high-frequency shaking in the near-field region raises concerns about the structural safety of buildings and facilities located near fracture networks, particularly in highly fractured geo-reservoirs close to population and industrial centers. Such fault networks may be activated by industrial activities, thus increasing the hazard.

With the availability of increasing computational resources, advanced numerical methods, and efficient modern codes, it has become feasible and affordable to integrate dynamic earthquake simulations into physics-based shaking assessments for seismic hazard analysis. However, this practice is not yet widespread. Adopting a rigorous physics-based approach that accounts for complex fault geometry, topography, and medium heterogeneity will help to significantly improve ground shaking estimates.

\subsection*{Limitations}
All results obtained from our simulations are contingent upon the model assumptions detailed in \citeA{palgunadi2024dynamic}. We acknowledge that specific findings, such as the spatial distribution of shaking or equivalent near-field characteristic frequency estimates, may vary with different configurations of fracture networks and fault models. Nonetheless, the fundamental physics underlying the dynamic interactions among fractures will remain consistent.

Our study primarily focuses on the geometrical complexity of faults while deliberately excluding several other factors. These include medium heterogeneities, low-velocity zones, milli-to-micro-scale fractures, topography, off-fault plastic deformation, fault roughness, site effects, and rupture dynamic interactions between faults and the free surface.

Another shortcoming of our study is the lack of observational data for spontaneous cascading fracture-network ruptures, which impedes more direct comparisons of our simulation outcomes with real earthquakes. A potential case study that may provide insights into such phenomena is the $\Mw 7.1$ 2019 Ridgecrest earthquake. However, the sparse distribution of seismic stations in the near-field region limits our capacity to investigate the signatures characteristic of fracture-network cascading earthquakes thoroughly. The search for the fracture-network cascading earthquake may be possible in a well-recorded moderate earthquake such as induced or triggered earthquakes in a geo-reservoir. 

\section*{Conclusions}
In this study, we investigate the near-source ground-motion properties of three high-resolution 3D dynamic rupture simulations within a geometrically complex fault zone featuring a main listric fault and over 800 multiscale fractures. The three analyzed cases include a cascading rupture within the fracture network, a main-fault non-cascading rupture with secondary slip in the fracture network, and a main-fault non-cascading rupture without a fracture network. 
Our results demonstrate that cascading ruptures within fracture networks generate high-frequency seismic radiation due to simultaneous slip on multiple fractures.
The incoherence of these dynamic rupture processes produces source coda wave-like signatures. The inferred high-resolution near-field characteristic frequency ($f_c$) and stress drop are significantly elevated in the fracture-network cascading rupture cases, approximately an order of magnitude higher than the estimation directly on the fault of the dynamic rupture simulation. In contrast, the non-cascading main-fault rupture scenarios show comparable $f_c$ and stress drop values. The fracture-network cascading ruptures show distinct azimuthal dependence of $f_c$ for epicentral distances $< 15$~km, with locally increased $f_c$ of vertical components reflecting the complex radiation pattern and rupture directivity. These results are consistent with observations from large-scale geometrically complex dynamic rupture simulations of real earthquakes and the well-recorded 2009–2016 Central Apennines, Italy, earthquake sequence. 
We identify several characteristics that highlight the unique ground-shaking patterns associated with cascading ruptures in fracture networks.
Simulated ground motions from cascading ruptures show consistently higher levels of peak ground acceleration (PGA), peak ground velocity (PGV), and pseudo-spectral acceleration (PSA) across azimuths, with PGA nearly double that of non-cascading scenarios. Our study emphasizes how fault zone complexities profoundly affect rupture dynamics and seismic wave radiation, and hence ground motion amplitude and distribution. High-performance computing enables detailed modeling of these effects, emphasizing the importance of incorporating geometrical complexity into physics-based seismic hazard assessment.

\section*{Data and Resources}\label{section:data_resources}
The version of SeisSol used in this study is described in \url{https://seissol.readthedocs.io/en/latest/fault-tagging.html#using-more-than-189-dynamic-rupture-tags} with commit version \verb|917250fd|. Patched meshing software PUMGen can be cloned from github branch PUMGenFaceIdentification64bit (\url{https://github.com/palgunadi1993/PUMGen/tree/PUMGenFaceIdentification64bit}). The strong motion data from three Central Italy's earthquakes can be found in \url{https://esm-db.eu/} (last access January, $10^{th}$ 2023) \cite{luzi2020engineering}. The AXITRA software can be downloaded in \url{https://github.com/coutanto/axitra/tree/master}. Various ground motion models are extracted from \url{https://github.com/bakerjw/GMMs}. All input and mesh files are available in the Zenodo repository at \citeA{palgunadi_2024_14363631}.

\section*{Declaration of Competing Interests}
The authors declare no competing interest.

\section*{Acknowledgement}
The authors thank the computational earthquake seismology (CES) group at KAUST for fruitful discussions and suggestions. The authors thank SeisSol's core developers (see \url{www.seissol.org}). Computing resources were provided by King Abdullah University of Science and Technology, Thuwal, Saudi Arabia (KAUST, Project k1587, k1488 and k1343 on Shaheen II; and Project k10043 and k10044 on Shaheen-3). The work presented in this article was supported by KAUST Grants (FRacture Activation in Geo-reservoir–physics of induced Earthquakes in complex fault Networks [FRAGEN], URF/1/3389-01-01, and BAS/1339-01-01). 
AAG acknowledges support by Horizon Europe (ChEESE-2P grant number 101093038, DT-GEO grant number 101058129, and Geo-INQUIRE grant number 101058518), the National Science Foundation (grants No. EAR-2121568, EAR-2121568,  OAC-2139536, OAC-2311208), the National Aeronautics and Space Administration (80NSSC20K0495), and the Southern California Earthquake Center (SCEC awards 22135, 23121). TU and AAG acknowledge support from the Bavarian State Ministry for Science and Art in the framework of the project Geothermal-Alliance Bavaria. DIG acknowledges support by the Natural Sciences and Engineering Research Council (Discovery Grant 05743). Part of the analysis was implemented using Obspy \cite{beyreuther2010obspy}. Figures were prepared using Paraview \cite{ahrens2005paraview} and Matplotlib \cite{hunter2007matplotlib}.

\bibliography{agusample.bib}

\section*{List of Figure Captions}

\newcommand{\textA}{Slip distribution in exploded view for cascading and non-cascading ruptures. This figure presents the slip distributions for two scenarios, depicted in an exploded view (expanded by a factor of 3.5 fracture-fracture distance): (a) cascading earthquake ($\Psi=65^\circ$) or \textit{Case 1}, and (b) non-cascading earthquake ($\Psi=120^\circ$) or \textit{Case 2}. The inset in the top right panel shows the original model configuration (non-exploded view) and the prestress ratio in the lower hemisphere projection. In the inset, white triangles and dots illustrate the polar representation of the fracture family 1 and 2, respectively. Whereas the grey dots denote the dip-dependence of the main fault surface. The black circle indicates the hypocenter. Notably, the non-cascading earthquake exhibits significantly higher peak slip values.}
\newcommand{\textB}{Schematic illustration of the seismogram generation workflow, to analyze fracture-specific radiation contributions. Opaque fracture planes represent the fractures actively slipping at $t=0.6$~s. Thin black lines illustrate individual waveform contributions from each slipped fracture, while the thick black lines depict the cumulative waveforms resulting from the summation of all individual contributions.}
\newcommand{\textC}{Comparison of seismograms obtained from a dynamic rupture simulation (solid red lines) with those generated through the combination of multiple kinematic simulations (dashed blue lines).  The root-mean-square (RMS) values indicate the discrepancies between the seismograms from the dynamic and kinematic simulations.}
\newcommand{\textD}{Waveforms and spectral analysis at selected stations for three scenarios with varying levels of activation of the main fault and embedded fractures: cascading rupture (\textit{Case 1}: fracture only, $\Psi=65^\circ$) in red, rupture with off-fault fracture slip (\textit{Case 2}: main fault + fractures, $\Psi=120^\circ$) in green, and rupture without off-fault fracture slip (\textit{Case 3}: main fault only, $\Psi=120^\circ$) in blue (after \citeA{palgunadi2024dynamic}. (a) Three-component (East-West; North-South, Vertical) waveform at selected stations (marked by red circles in the inset figure showing station distribution). 
(b) Fourier velocity spectra [m/s Hz$^{-1}$]. The grey vertical dashed line marks an estimate of the highest resolved frequency ($f \sim 7$ Hz) for these stations in the dynamic rupture model.}
\newcommand{\textE}{Azimuthal dependence of equivalent near-field characteristic frequency ($f_c$) for \textit{Case 1} (cascading rupture within the fracture network). This figure examines stations within a 4.8 and 5.2 km epicentral distance using the vertical component.
(a) Three examples of $f_c$ determination using Brune's model (solid red line) from seismogram displacement spectrum. The grey rectangle highlights the estimated confidence interval for $f_c$, while the blue rectangle indicates the selected $f_c$ value. 
(b) Corner frequency versus seismic moment ($M_0$), comparing $f_c$ estimates from \citeA{allmann2009global} for moderate to large magnitude earthquakes (white dots), with our estimates for \textit{Case 1}, \textit{Case 2} (rupture with off-fault fracture slip), and \textit{Case 3} (main fault rupture without the fracture network) (grey circle, triangle, and square, and error bars). Dashed red lines indicate constant stress drops of 0.1, 1, 10, and 100 MPa. 
(c) Azimuthal dependence of $f_c$ and associated error bars for \textit{Case 1}. The red dashed line represents the mean value of $f_c$.}
\newcommand{\textF}{Corner frequency variation with azimuth for the 2016 $\Mw$ 6.6 Norcia earthquake. This figure illustrates how the corner frequency may vary relative to the azimuth from the epicenter in natural earthquakes. The fault-plane solution, depicting the focal mechanism, is displayed in the top left corner.}
\newcommand{\textG}{Spatial variation of equivalent near-field characteristic frequency ($f_c$) across different components. $f_c$ are calculated from the dynamic rupture simulation's seismograms (\textit{Case 1}: cascading rupture within the fracture network) based on the vertical, East-West, and North-South components, respectively. The white and black dashed lines indicate the radiation pattern of SH and SV waves, respectively, which allows interpreting parts of the $f_c$ variations. The vertical component inferred $f_c$ shows patterns that reflect the  SV-wave radiation and the strike orientation of the two fracture families. The East-West component (main-fault parallel) inferred $f_c$ are asymmetrically distributed and appear correlated with SH-wave radiation. The North-South component (main-fault normal) inferred $f_c$ displays 4 marked narrow lobes of higher corner frequencies. The black circles mark the region analyzed in Figure \ref{fig:FcAzimuth}. Black line represents the top of the main fault and it is dipping to the North.}
\newcommand{\textH}{Comparison of simulated Peak Ground Acceleration (PGA), panel (a) and Peak Ground Velocity (PGV), panel (b) against a selection of ground motion models (GMMs). We apply a site-amplification correction factor \cite{borcherdt1994estimates, borcherdt2002empirical} to the simulated waveforms, based on homogeneous elastic properties, with S-wave speed of 3.464 km/s, and assume $V_{S30}$ = 0.76 km/s for the site-amplification correction. Thick colored lines represent GMMs for $\Mw=5.5$, and dashed colored lines indicate GMMs for $\Mw=6.0$. The simulation results are expressed as GMRotD50, a metric independent of the sensor orientations. Red dots, green triangles, and blue squares depict PGA and PGV values obtained from the cascading rupture, rupture with off-fault fracture slip, and rupture without off-fault fracture slip, respectively.
\textit{Case 1} is the cascading rupture within the fracture network, \textit{Case 2} is the rupture with off-fault fracture slip, and \textit{Case 3} is the main fault rupture without the fracture network.}
\newcommand{\textI}{Azimuths and distance dependence of Peak Ground Acceleration (PGA) and Peak Ground Velocity (PGV). PGA and PGV are plotted against the Joyner-Boore distance ($R_{JB}$) for different azimuths and for \textit{Cases 1}, \textit{2}, and \textit{3}, and compared with selected ground motion models (GMMs) from \citeA{abrahamson2014summary, boore2014nga, campbell2014nga, chiou2014update}. We use GMRotD50 as well as a site-amplification correction factor \cite{borcherdt1994estimates, borcherdt2002empirical} to correct for the homogeneous S-wave speed (3.464 km/s) in the simulations with respect to the assumed $V_{S30}$ = 0.76 km/s in the GMMs. \textit{Case 1} represents a cascading rupture within the fracture network. \textit{Case 2} corresponds to a rupture with off-fault fracture slip. \textit{Case 3} is a main fault rupture without the fracture network. Panels (a) and (d) depict PGA and PGV for \textit{Case 1}, respectively, with solid lines showing GMM predictions for $\Mw = 5.5$. Panels (b) and (e) illustrate PGA and PGV for \textit{Case 2}, respectively, with GMM predictions based on $\Mw = 6.0$. Panels (c) and (f) display PGA and PGV for \textit{Case 3}, respectively, also with GMM predictions based on $\Mw = 6.0$.}
\newcommand{\textJ}{Shakemaps of PGA and PGV. This figure presents PGA (top row) and PGV (bottom row), measured in $g$ and $m/s$ respectively, using GMRotD50 \cite{boore2006orientation}. Panels (a) and (d) depict results for \textit{Case 1} (cascading rupture within the fracture network), panels (b) and (e) for \textit{Case 2} (rupture with off-fault fracture slip), and panels (c) and (f) for \textit{Case 3} (main fault rupture without the fracture network). In each panel, the rectangle outlines the main listric fault plane, and the thick black line indicates its top. The white star marks the epicenter.}
\newcommand{\textK}{Azimuthal dependence of ground-motion intensities. This figure illustrates the variation in (a) Peak Ground Acceleration (PGA), (b) Peak Ground Velocity (PGV), and (c) Pseudo Spectral Acceleration (PSA) at T=0.3~s (SA(T=0.3~s)) across different azimuths, showing average values and standard deviations. The red, blue, and green lines correspond to the cascading rupture (\textit{Case 1}), rupture with off-fault fracture slip (\textit{Case 2}), and rupture without off-fault fracture network in the model (\textit{Case 3}), respectively. Transparent colored boxes overlay the figure to denote the strike orientation of fracture families 1 and 2, with each box's width representing a $10^\circ$ standard deviation from the average strike orientation of the fractures.}
\newcommand{\textL}{Comparison of pseudo-spectral acceleration (PSA) with ground motion models (GMMs). This figure compares PSA values from cascading and non-cascading earthquake scenarios against selected GMMs \cite{abrahamson1997empirical, boore1997equations, boore2014nga, campbell2014nga}. \textit{Case 1} represents a cascading rupture within the fracture network, \textit{Case 2} corresponds to a rupture with off-fault fracture slip, and \textit{Case 3} is a main fault rupture without the fracture network. Panel (a) depicts PSA (damping $\zeta$ = 0.05) versus period at a distance of 10 km from the hypocenter for cascading rupture (red triangles), rupture with off-fault fracture slip (green circles), and rupture without off-fault fracture slip (blue squares). Thick lines indicate GMMs for $\Mw=5.5$ and dashed lines for $\Mw=6.0$. Panel (b) shows PSA at $T$ = 0.3~s, panel (c) at $T$ = 1~s, and panel (d) at $T$ = 2~s, each plotted against the Joyner-Boore distance ($R_{JB}$) for the three cases.}
\newcommand{\textM}{Map of Spectral Acceleration for $T$ = 0.3~s. This map shows spectral acceleration (in $g$) across three scenarios: (a) \textit{Case 1}: cascading rupture within the fracture network, (b) \textit{Case 2}: rupture with off-fault fracture slip, and (c) \textit{Case 3}: main fault rupture without the fracture network in the model. The rectangle outlines the listric fault plane, with a thicker line denoting the top of the main fault. The white star marks the epicenter.}

\begin{figure}
    \centering
    \includegraphics[width=0.99\textwidth]{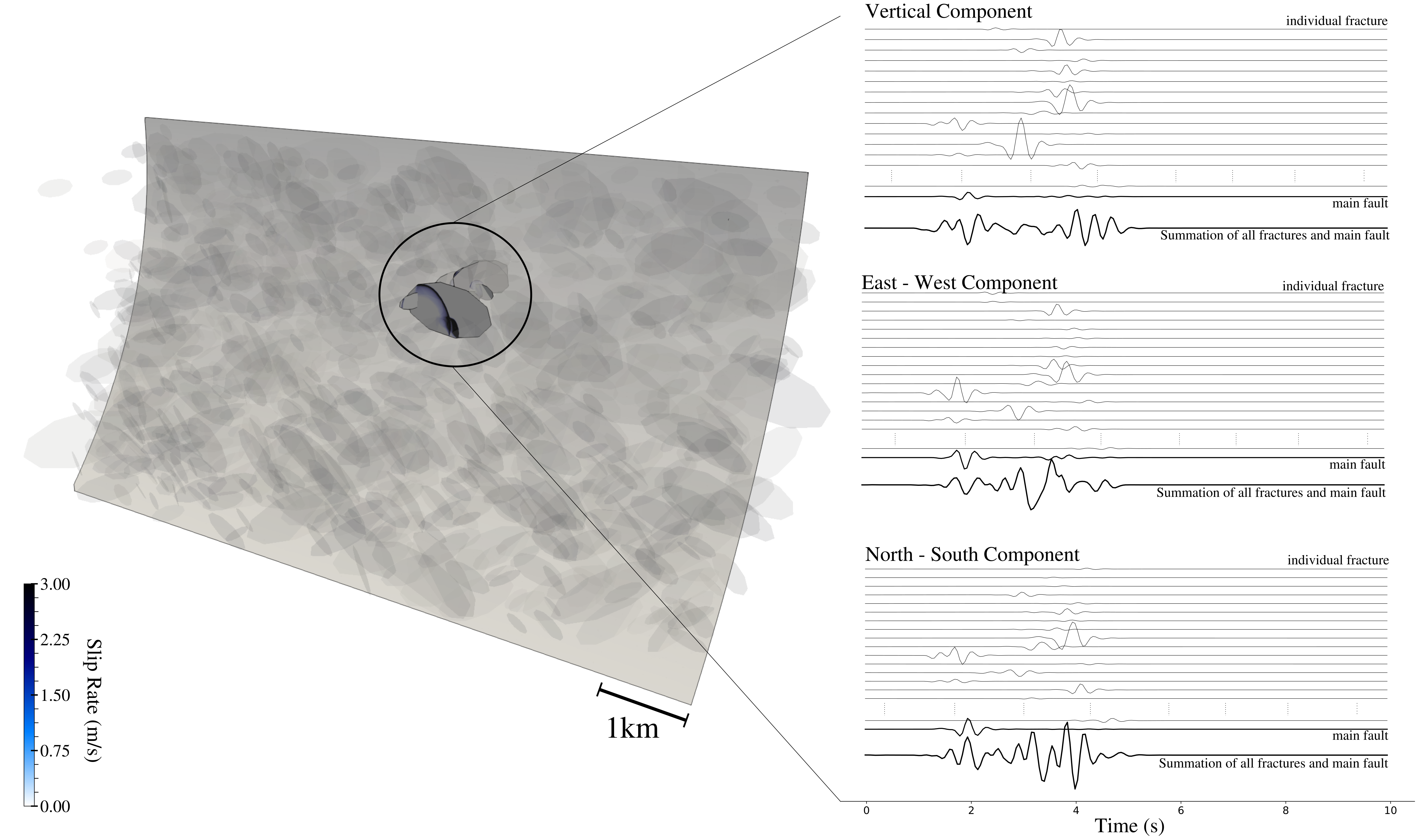}
    \caption{\textB}
    \label{fig:Schematic}
\end{figure}

\begin{figure}
    \centering
    \includegraphics[width=0.7\textwidth]{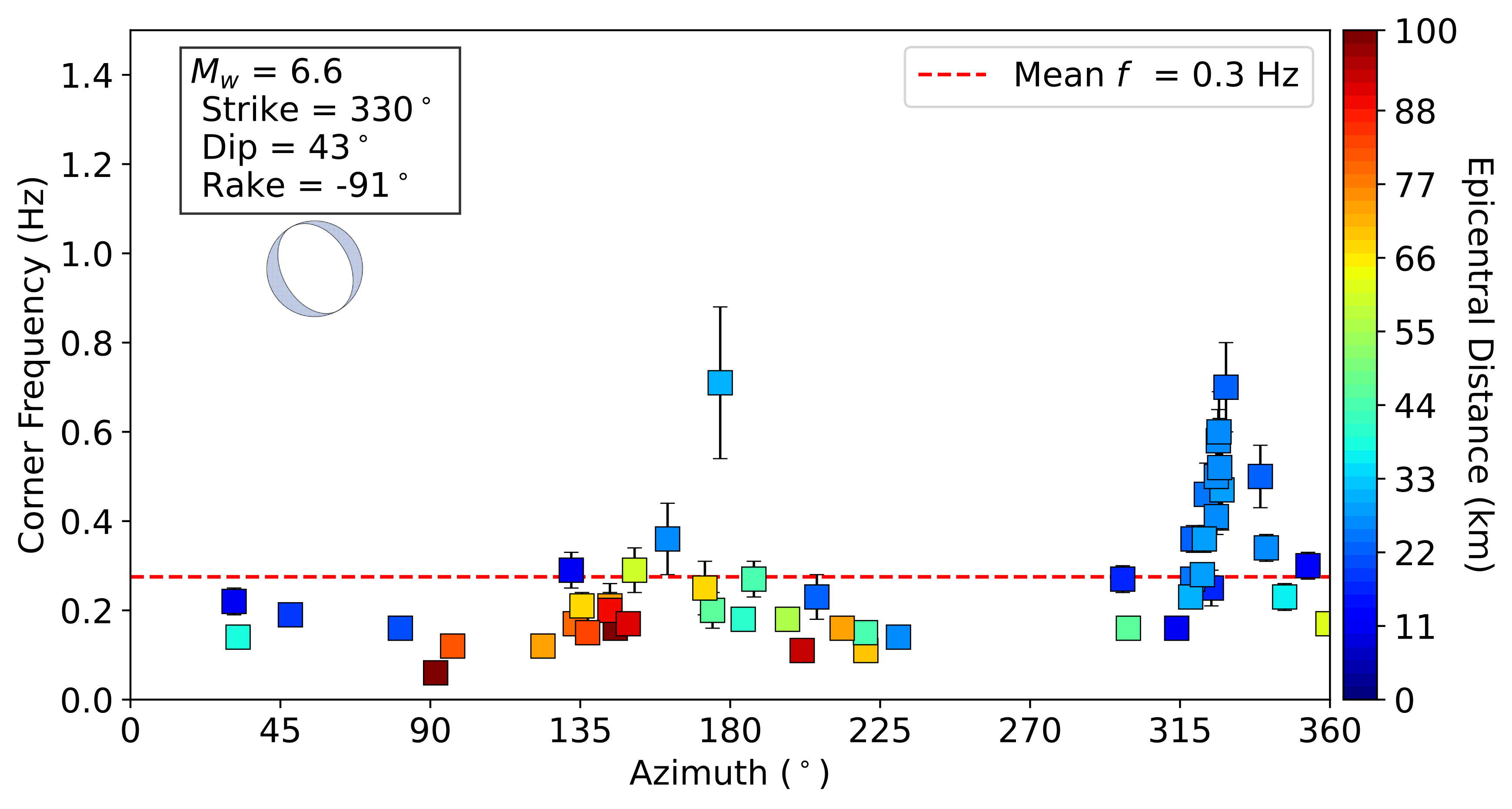}
    \caption{\textF}
    \label{fig:fc_norcia}
\end{figure}

\begin{figure}
    \centering
    \includegraphics[width=0.99\textwidth]{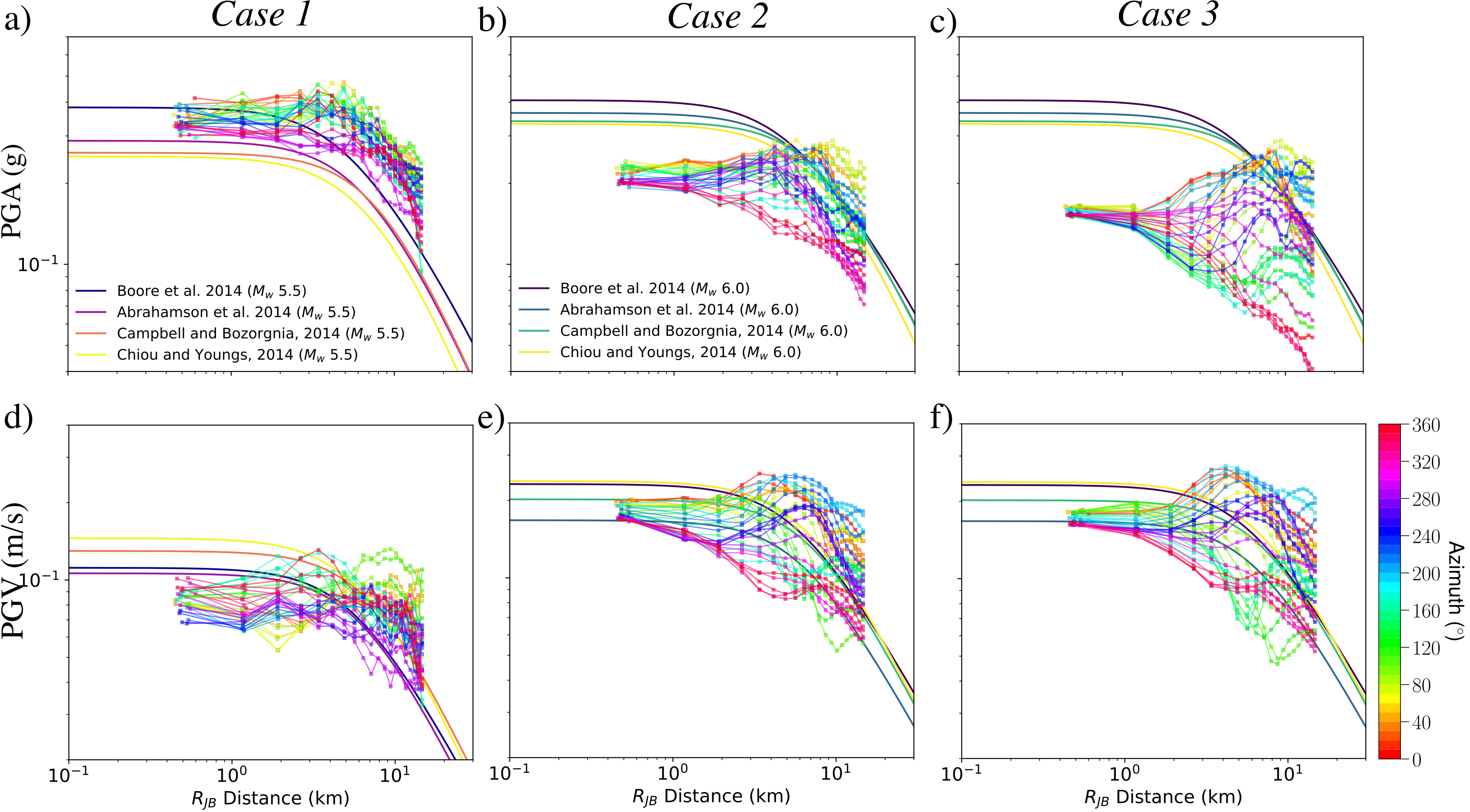}
    \caption{\textI}
    \label{fig:pga_pgv_azimuth}
\end{figure}

\begin{figure}
    \centering
    \includegraphics[width=0.99\textwidth]{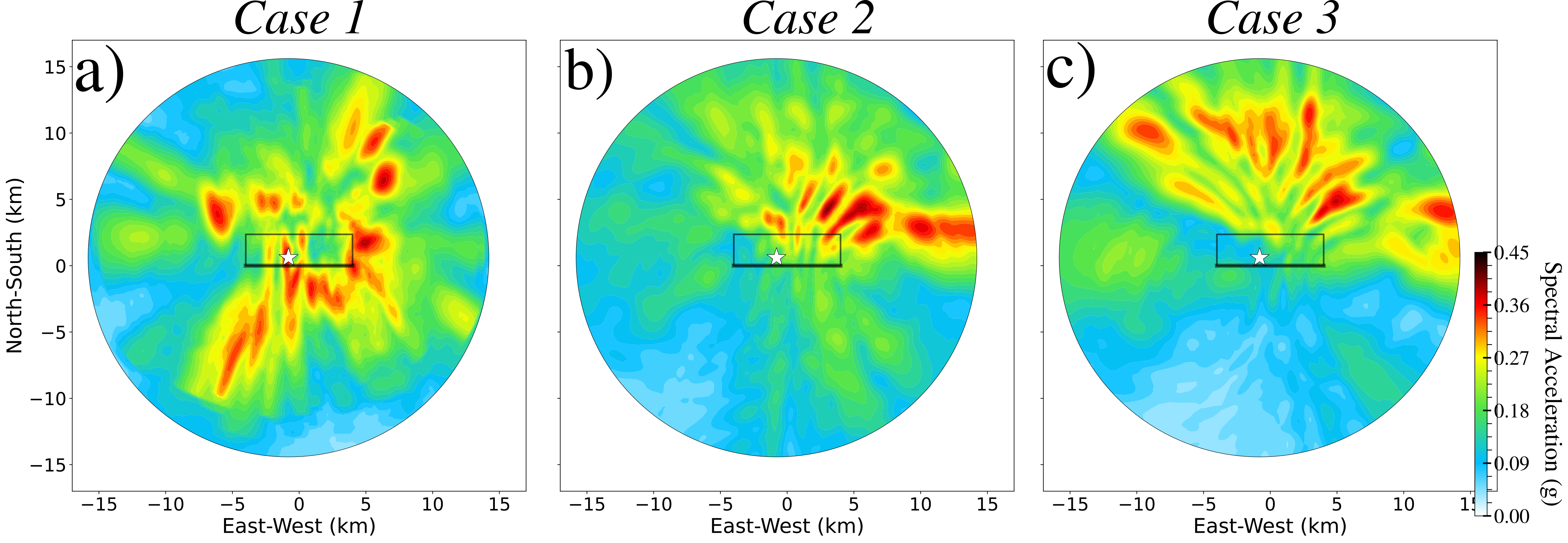}
    \caption{\textM}
    \label{fig:SAmap}
\end{figure}

\begin{itemize}
    \item Figure \ref{fig:slip}: \textA
    \item Figure \ref{fig:Schematic}: \textB
    \item Figure \ref{fig:CompareSeismogram}: \textC
    \item Figure \ref{fig:WaveformSpectral}: \textD
    \item Figure \ref{fig:FcAzimuth}: \textE
    \item Figure \ref{fig:fc_norcia}: \textF
    \item Figure \ref{fig:CornerFreq}: \textG
    \item Figure \ref{fig:GMM}: \textH
    \item Figure \ref{fig:pga_pgv_azimuth}: \textI
    \item Figure \ref{fig:PGA}: \textJ
    \item Figure \ref{fig:azimuthal_groundMotions}: \textK
    \item Figure \ref{fig:SAgraph}: \textL
    \item Figure \ref{fig:SAmap}: \textM
\end{itemize}

\clearpage
\section*{Figures}

\begin{figure}[htp!]
    \centering
    \includegraphics[width=0.95\textwidth]{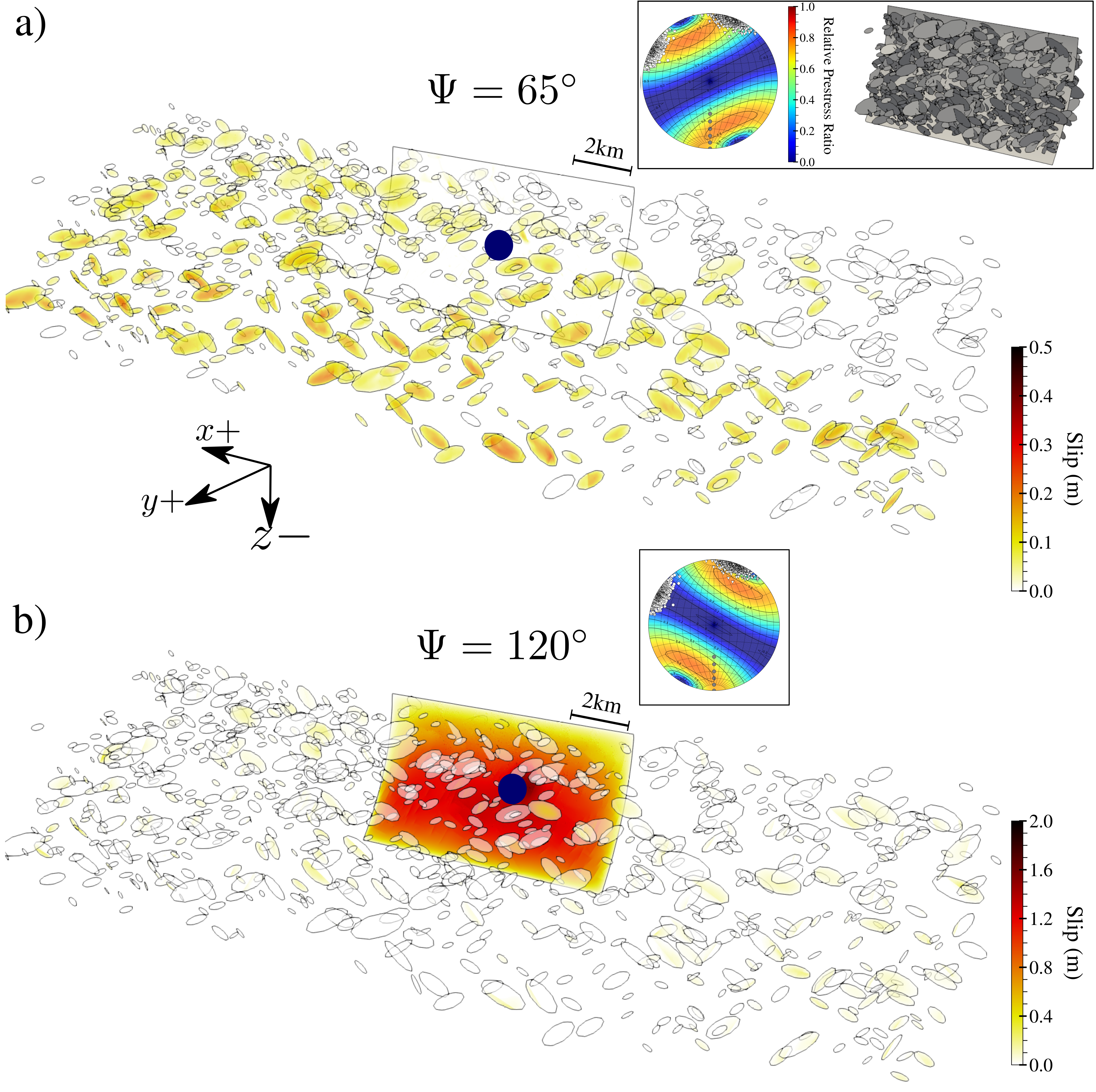}
    \caption{\textA}
    \label{fig:slip}
\end{figure}

\begin{figure}
    \centering
    \includegraphics[width=0.99\textwidth]{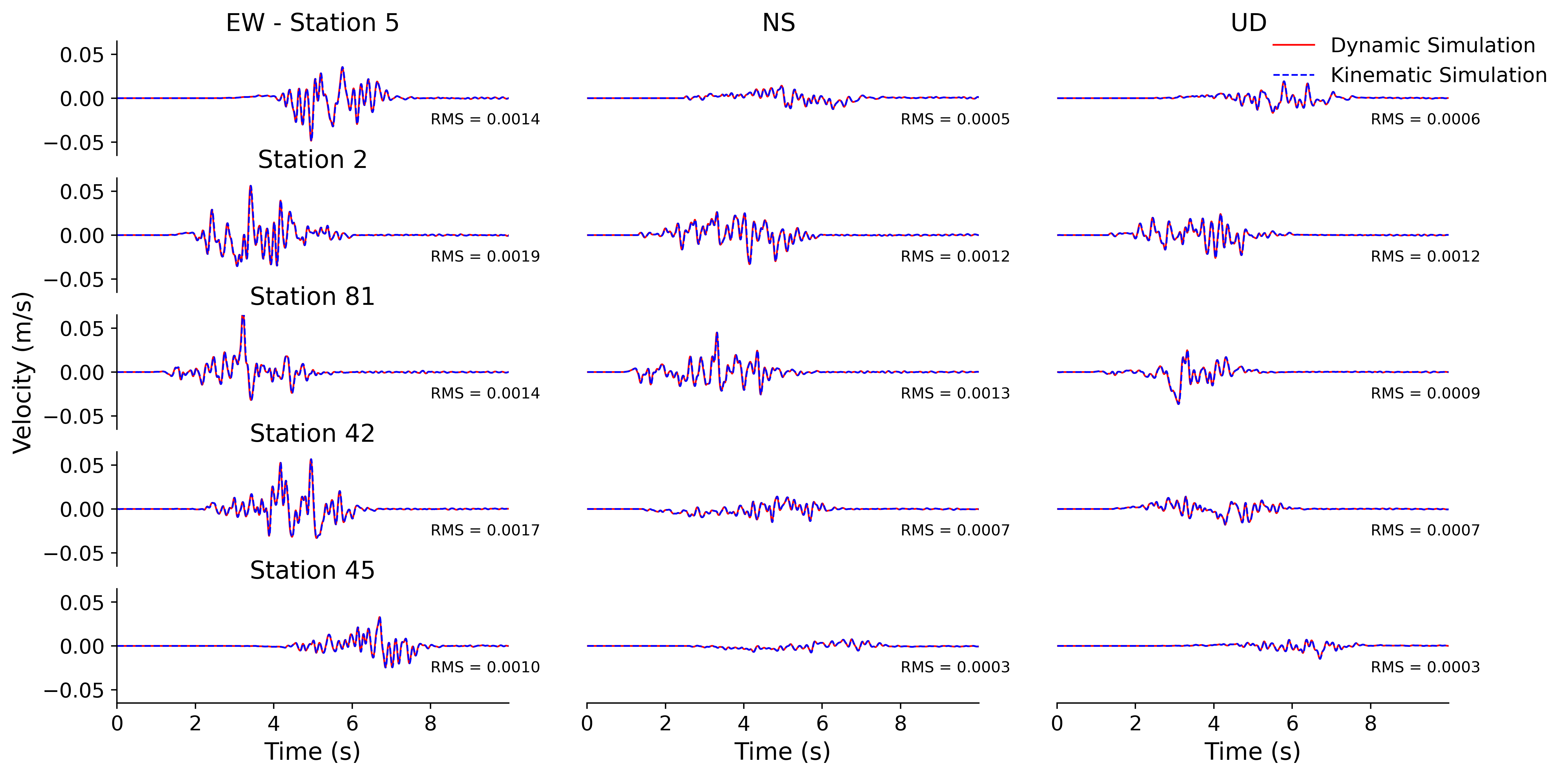}
    \caption{\textC}
    \label{fig:CompareSeismogram}
\end{figure}

\begin{figure}
    \centering
    \includegraphics[width=0.95\textwidth]{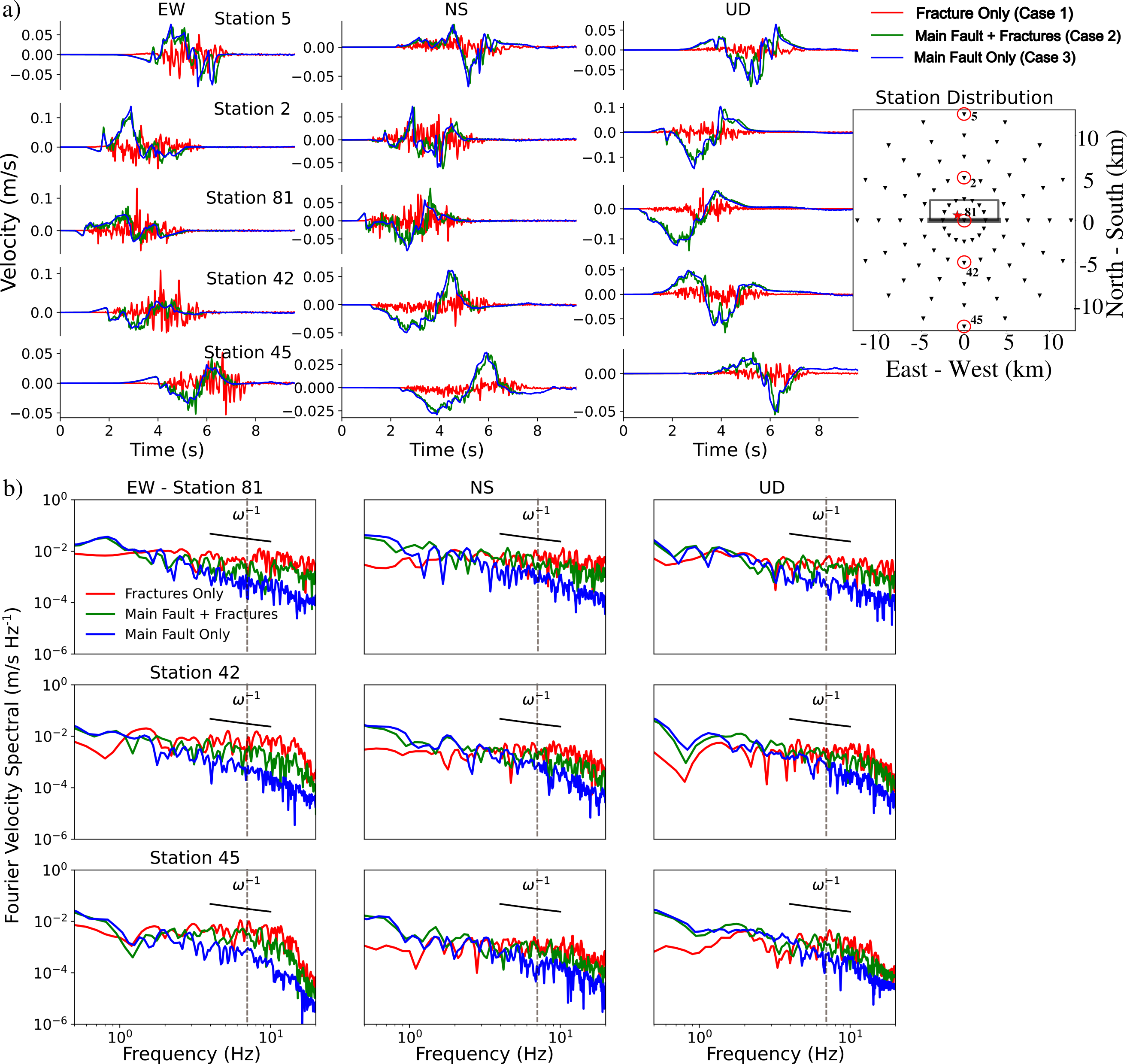}
    \caption{\textD}
    \label{fig:WaveformSpectral}
\end{figure}

\begin{figure}
    \centering
    \includegraphics[width=0.9\textwidth]{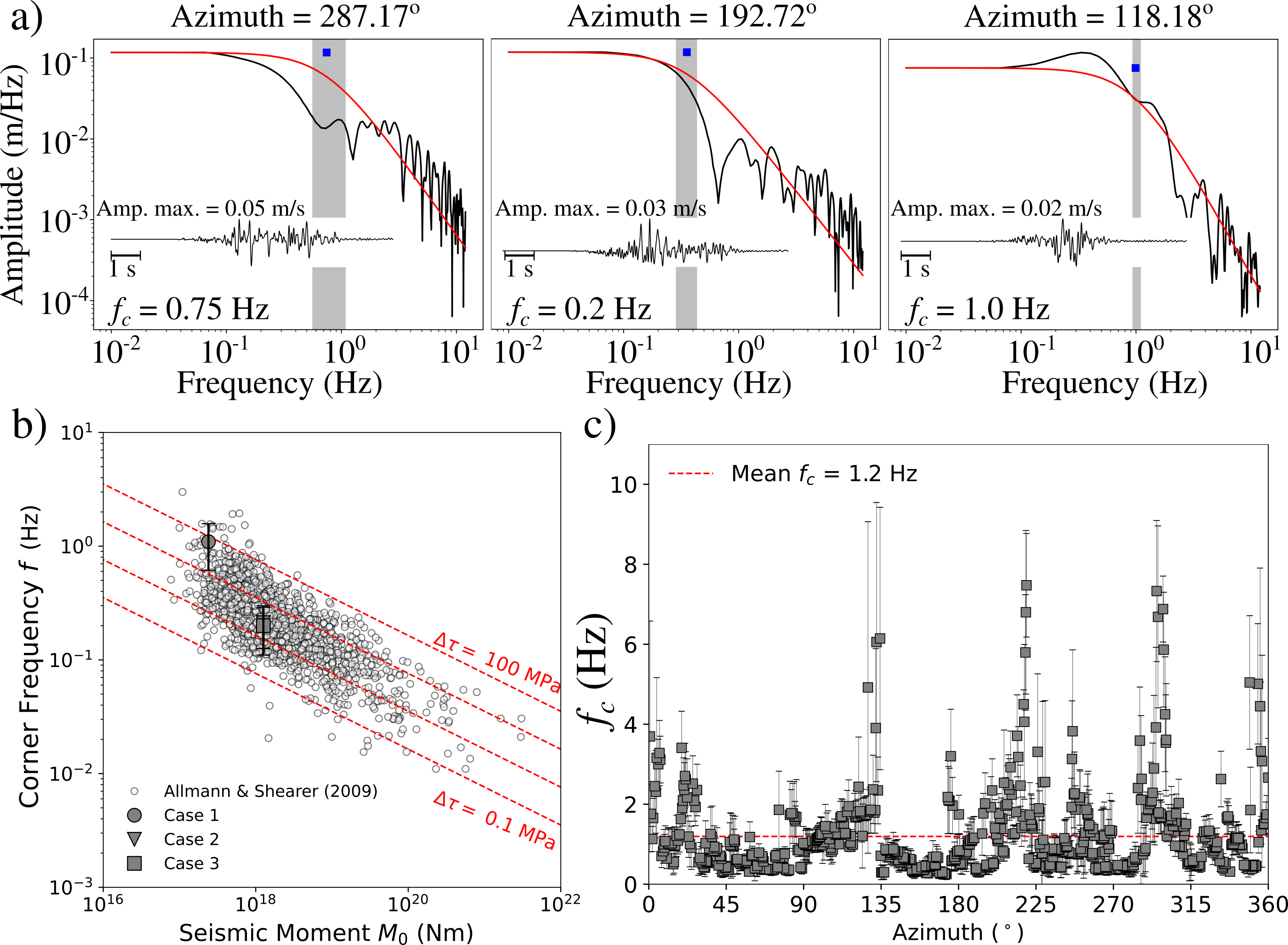}
    \caption{\textE}
    \label{fig:FcAzimuth}
\end{figure}

\begin{figure}
    \centering
    \includegraphics[width=0.99\textwidth]{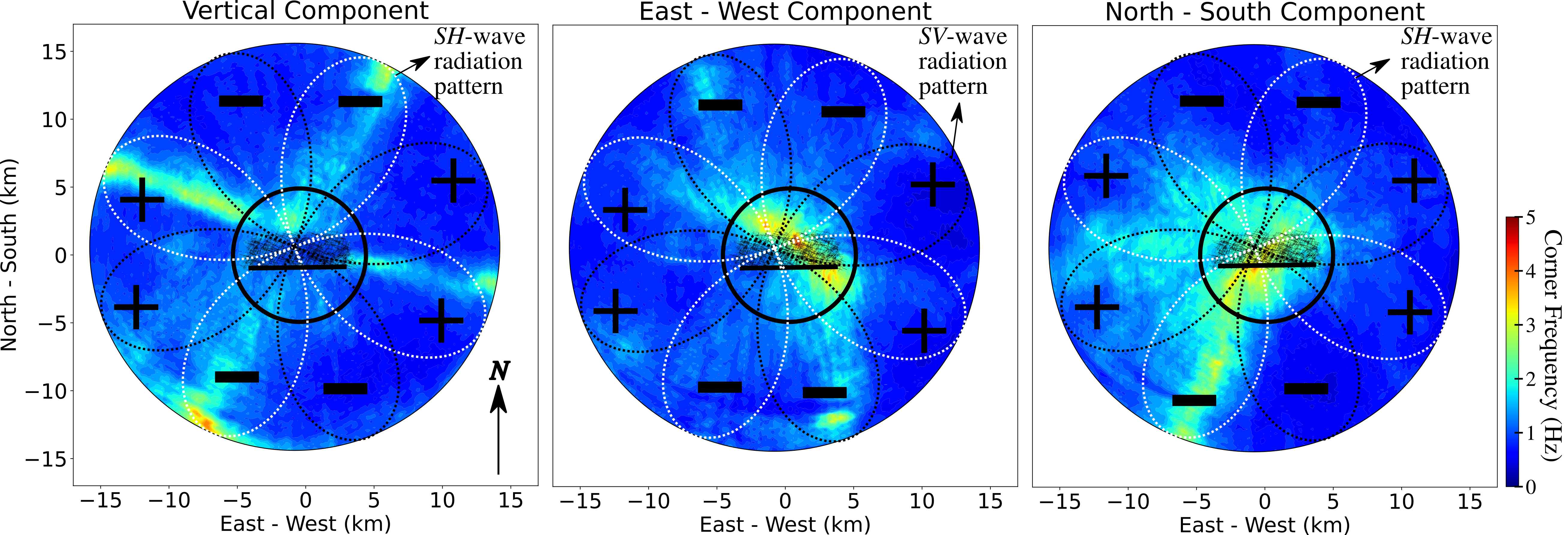}
    \caption{\textG}
    \label{fig:CornerFreq}
\end{figure}

\begin{figure}
    \centering
    \includegraphics[width=0.95\textwidth]{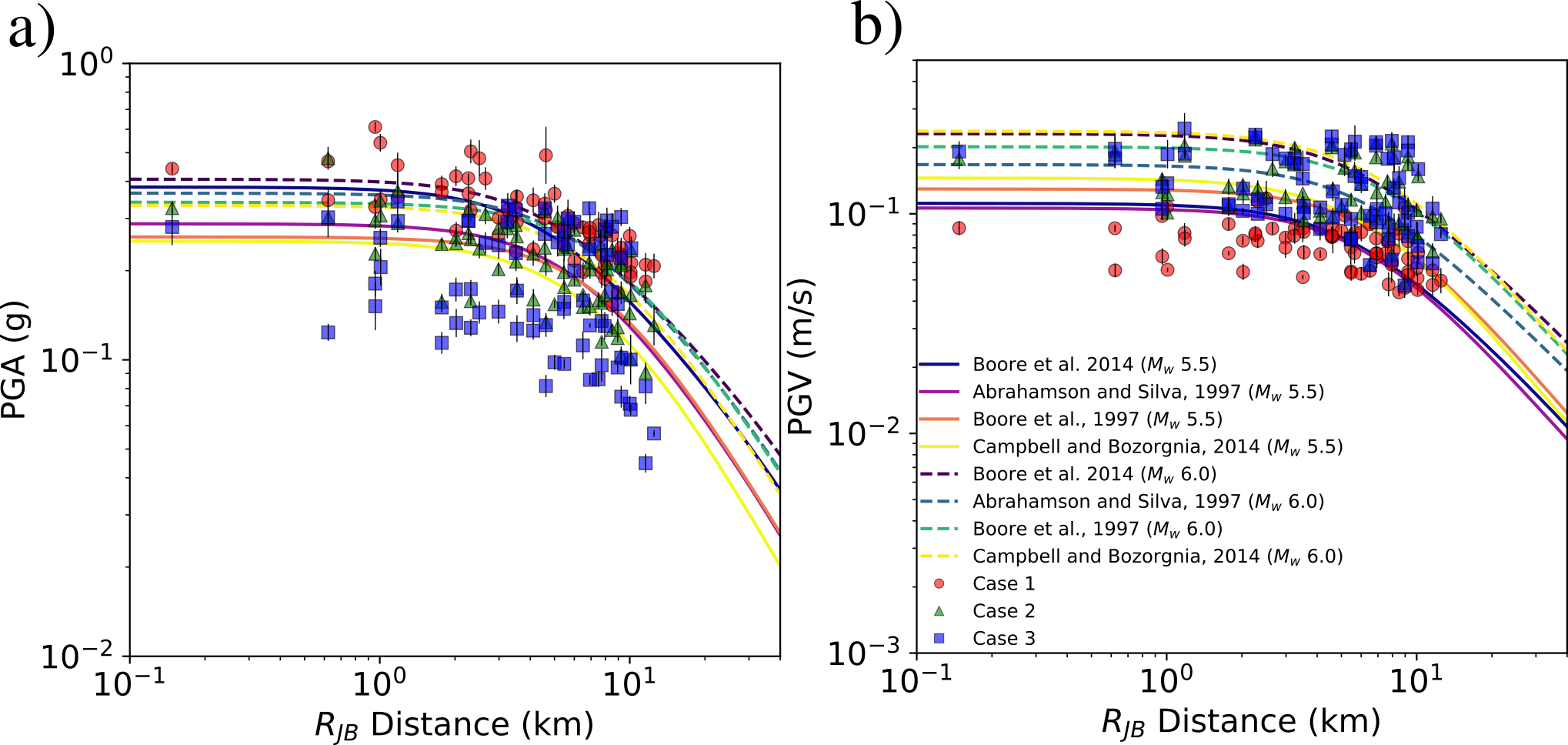}
    \caption{\textH}
    \label{fig:GMM}
\end{figure}

\begin{figure}
    \centering
    \includegraphics[width=0.99\textwidth]{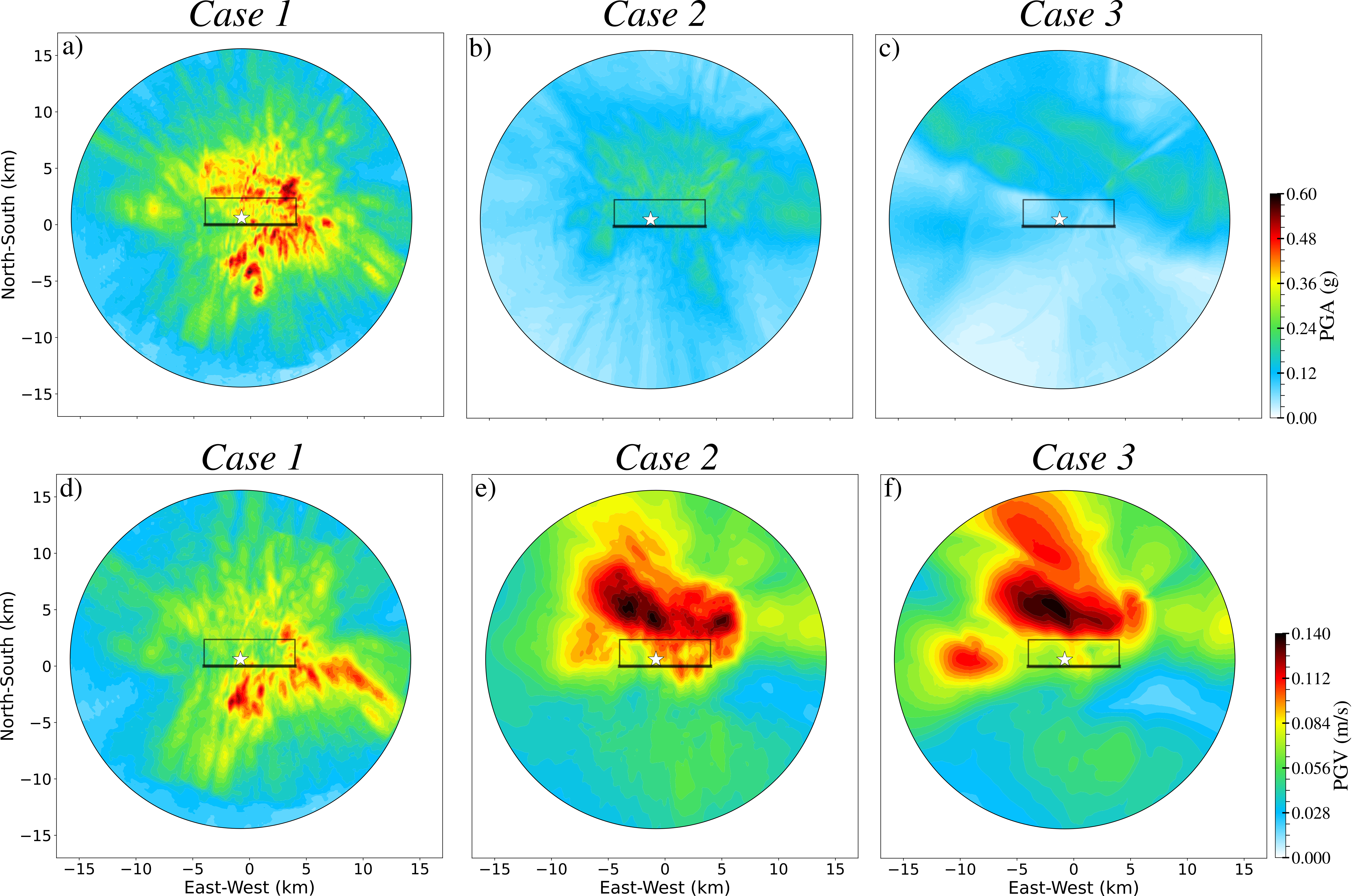}
    \caption{\textJ}
    \label{fig:PGA}
\end{figure}

\begin{figure}
    \centering
    \includegraphics[width=0.99\textwidth]{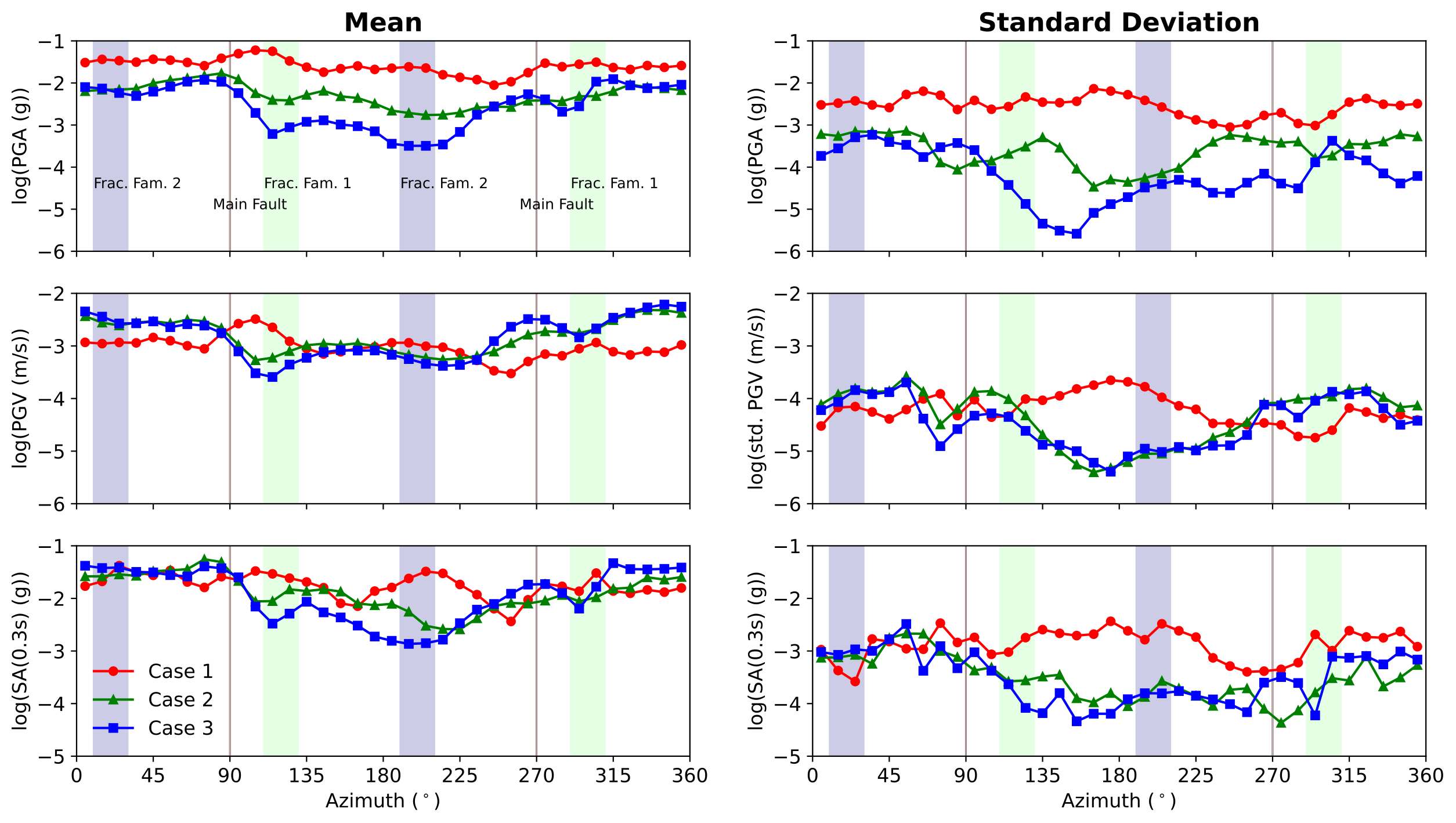}
    \caption{\textK}
    \label{fig:azimuthal_groundMotions}
\end{figure}

\begin{figure}
    \centering
    \includegraphics[width=0.95\textwidth]{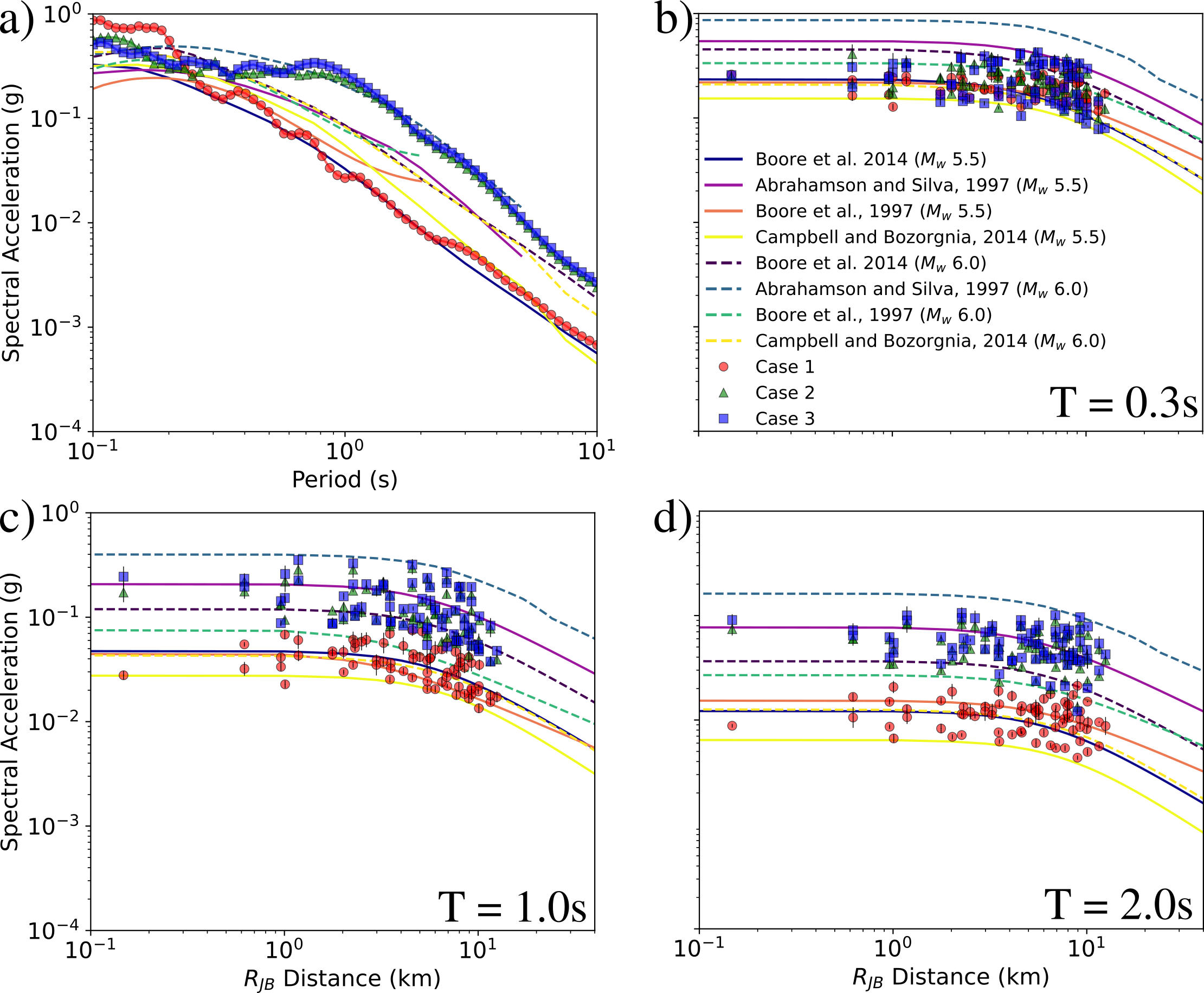}
    \caption{\textL}
    \label{fig:SAgraph}
\end{figure}

\clearpage
 \setcounter{figure}{0}
 \renewcommand{\thefigure}{S\arabic{figure}}

\title{Supporting Information for "Ground Motion Characteristics of Cascading Earthquake in a Multiscale Fracture Network"}

\authors{Kadek Hendrawan Palgunadi\affil{1,\dagger}, Alice-Agnes Gabriel\affil{2,4}, Dmitry Igor Garagash\affil{3}, Thomas Ulrich\affil{4},Nico Schliwa\affil{4}, Paul Martin Mai\affil{1}}

\affiliation{1}{Physical Science and Engineering, King Abdullah University of Science and Technology, Thuwal, Saudi Arabia}
\affiliation{2}{Institute of Geophysics and Planetary Physics, Scripps Institution of Oceanography, University of California, San Diego, CA, USA}
\affiliation{3}{Dalhousie University, Department Civil Resource Engineering, Halifax, Canada}
\affiliation{4}{Department of Earth and Environmental Sciences, Geophysics, Ludwig-Maximilians-Universit\"{a}t M\"{u}nchen, Munich, Germany}
\affiliation{\dagger}{now at Swiss Seismological Service (SED), ETH Zürich, Switzerland; Department of Geophysical Engineering, Institut Teknologi Sepuluh Nopember, Indonesia.}

\noindent\rule{\textwidth}{1pt}

\section*{Contents of this file:}
\begin{enumerate}
    \item Figure \ref{fig:model}: The 3D numerical model.
    \item Figure \ref{fig:riseTime}: Rise time for the two scenarios. 
    \item Figure \ref{fig:FcDistributionEW}: Azimuthal-dependence of corner frequency for EW component.
    \item Figure \ref{fig:FcDistributionNS}: Azimuthal-dependence of corner frequency for NS component.
    \item Figure \ref{fig:fc_statistics}: Corner frequency's statistics.
    \item Figure \ref{fig:fc_realEarthquake}: Azimuthal-dependence of corner frequency for real earthquakes.
    \item Figure \ref{fig:corrections}: Waveform deamplification using the site-amplification factors.
    \item Figure \ref{fig:rupture_speed_statistics}: Histogram of the rupture speed.
\end{enumerate}

\section*{Summary}
\begin{figure}[htp!]
    \centering
    \includegraphics[width=0.9\textwidth]{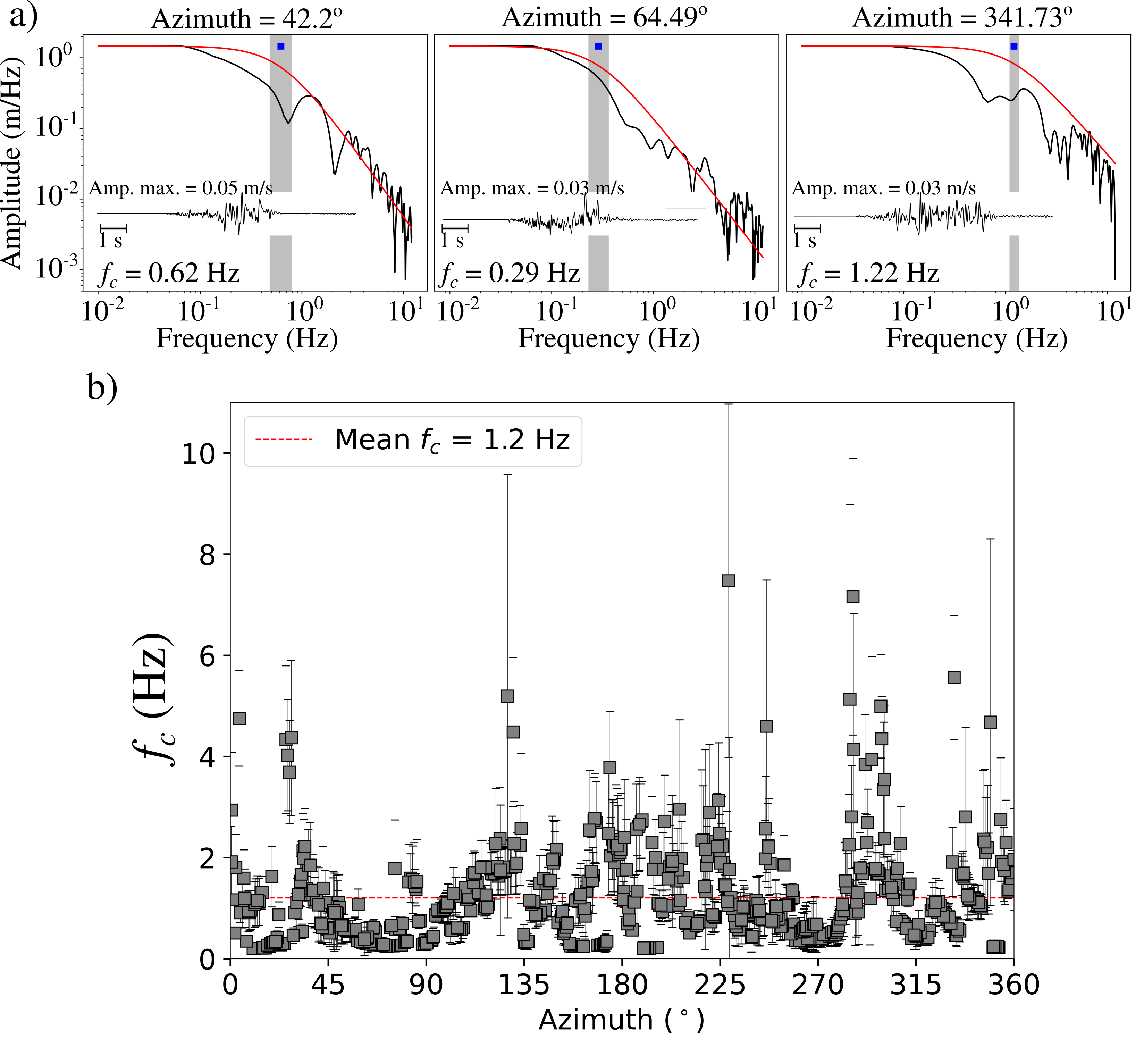}
    \caption{Azimuthal dependence of the equivalent near-field characteristic frequency ($f_c$) for the cascading earthquake on stations within a 4.8 and 5.2 km, considering the East-West component. The figure complements Figure 5, based on the vertical component. a) Three examples of $f_c$ determination using the Brune model (solid red line) from their corresponding seismograms. The grey rectangle represents the error range of $f_c$, while the blue rectangle denotes the selected $f_c$ value. b) Azimuthal dependence of $f_c$. The line inside the grey square represents the error range of $f_c$, and the red dashed line indicates the mean value of $f_c$.}
    \label{fig:FcDistributionEW}
\end{figure}

\begin{figure}[htp!]
    \centering
    \includegraphics[width=0.9\textwidth]{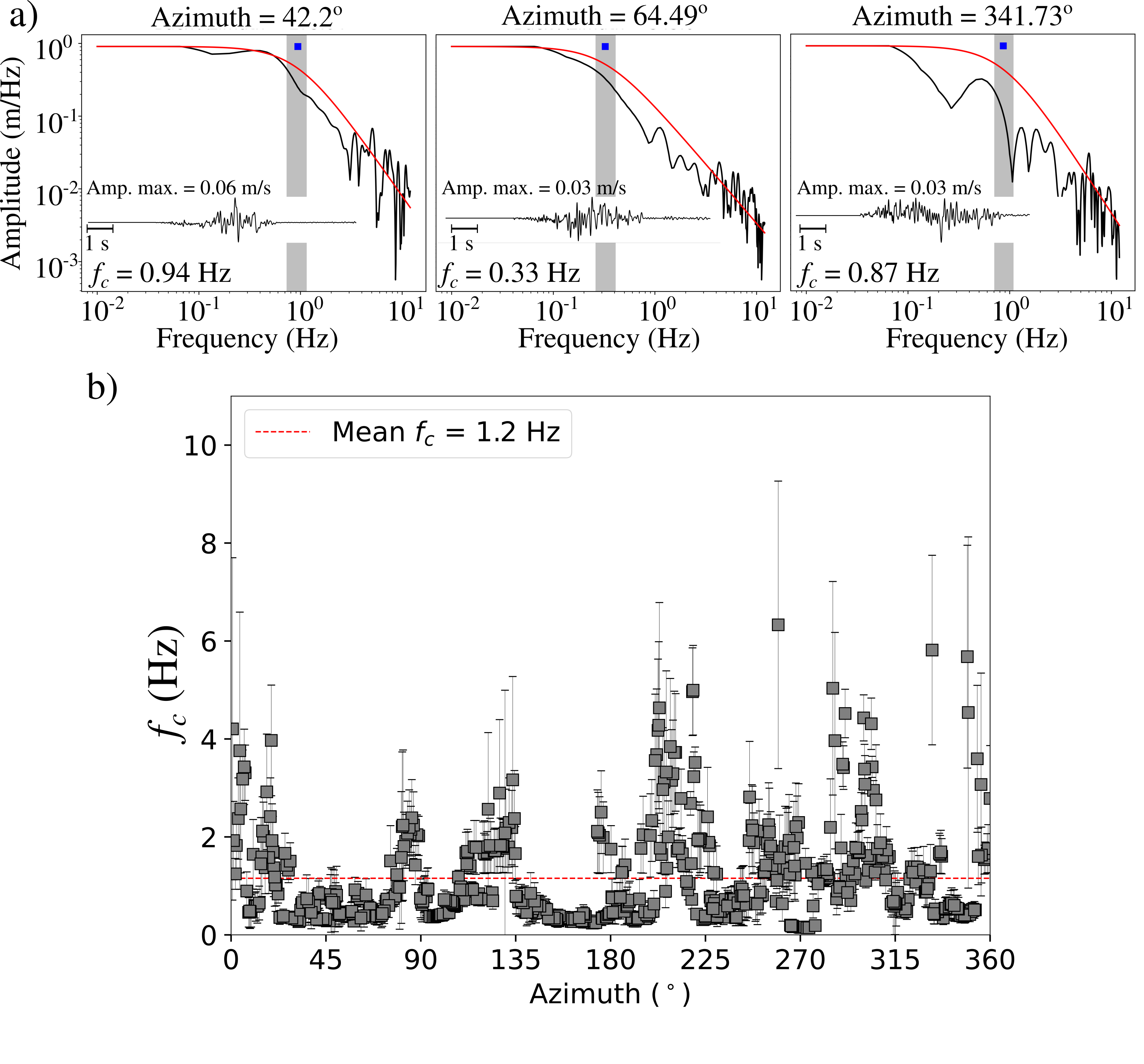}
    \caption{Same as Figure \ref{fig:FcDistributionEW} but for North-South component.}
    \label{fig:FcDistributionNS}
\end{figure}

The supporting figures depict the 3D model used in this study (Figure \ref{fig:model}), rise time of the two scenarios described in the main text in Figure \ref{fig:riseTime} and the azimuthal-dependence of the equivalent near-field characteristic frequency ($f_c$) for the cascading rupture in the East-West (EW) and North-South (NS) components (Figure \ref{fig:FcDistributionEW} and \ref{fig:FcDistributionNS}). Figure \ref{fig:fc_statistics} histograms of corner-frequency measurements for different cases in this study. Figure \ref{fig:fc_realEarthquake} displays the azimuthal dependence of $f_c$ for two recent well-recorded earthquakes. Figure \ref{fig:corrections} provides an example of the effects of applying a site-amplification correction on seismograms on one station before comparing it to ground motion models. Figure \ref{fig:rupture_speed_statistics} shows the histogram of the rupture speed from two cases.

\section*{Figures:}

\begin{figure}
    \centering
    \includegraphics[width=0.9\linewidth]{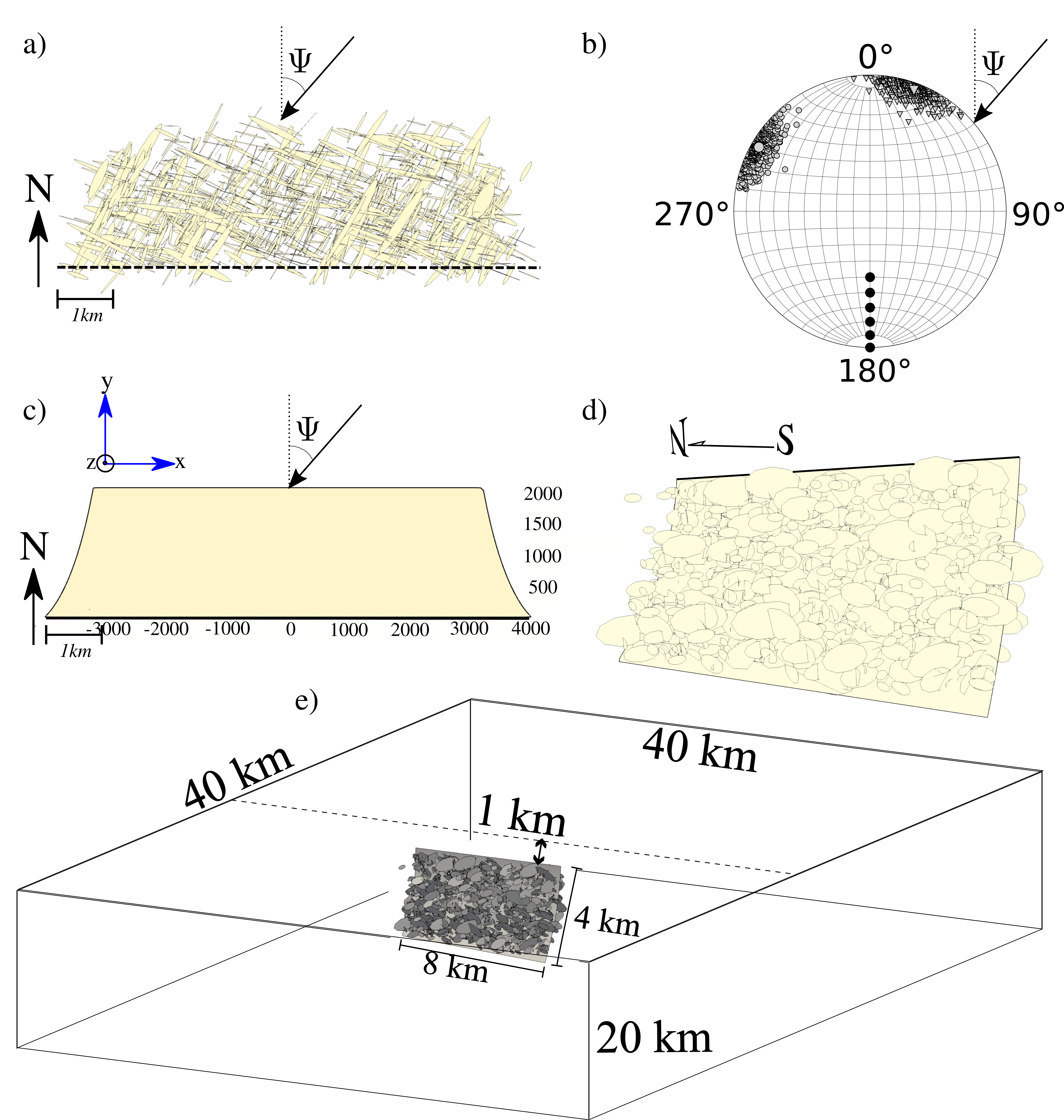}
    \caption{\small Geometric configuration of the 3D dynamic rupture model, incorporating a listric main fault and over 800 multiscale fractures, reproduced from Figure 2 of \citeA{palgunadi2024dynamic}. (a) Map view displaying the spatial distribution of the fracture network. The black dashed line indicates the shallow top of the main fault, while the black arrow represents the direction ($\Psi$) of the maximum horizontal stress, measured clockwise from North. (b) Polar view of fractures and the main fault presented in a lower hemisphere projection, aligned with the local fault normal orientation. Gray dots and triangles denote two distinct fracture families, averaging strikes of N120°E (dip 84°, gray triangles) and N20°E (dip 84°, large gray dots). Black dots illustrate the depth-dependent dip of the listric fault. (c) Map view of the listric fault, with the thick black line marking the fault’s top surface. (d) Perspective view of the listric fault geometry surrounded by 854 multiscale fractures, viewed from the east-southeast (ESE) direction. The thick black line highlights the top of the listric fault. (e) Representation of the 3D numerical model domain.}
    \label{fig:model}
\end{figure}

\begin{figure}[htp!]
    \centering
    \includegraphics[width=0.95\textwidth]{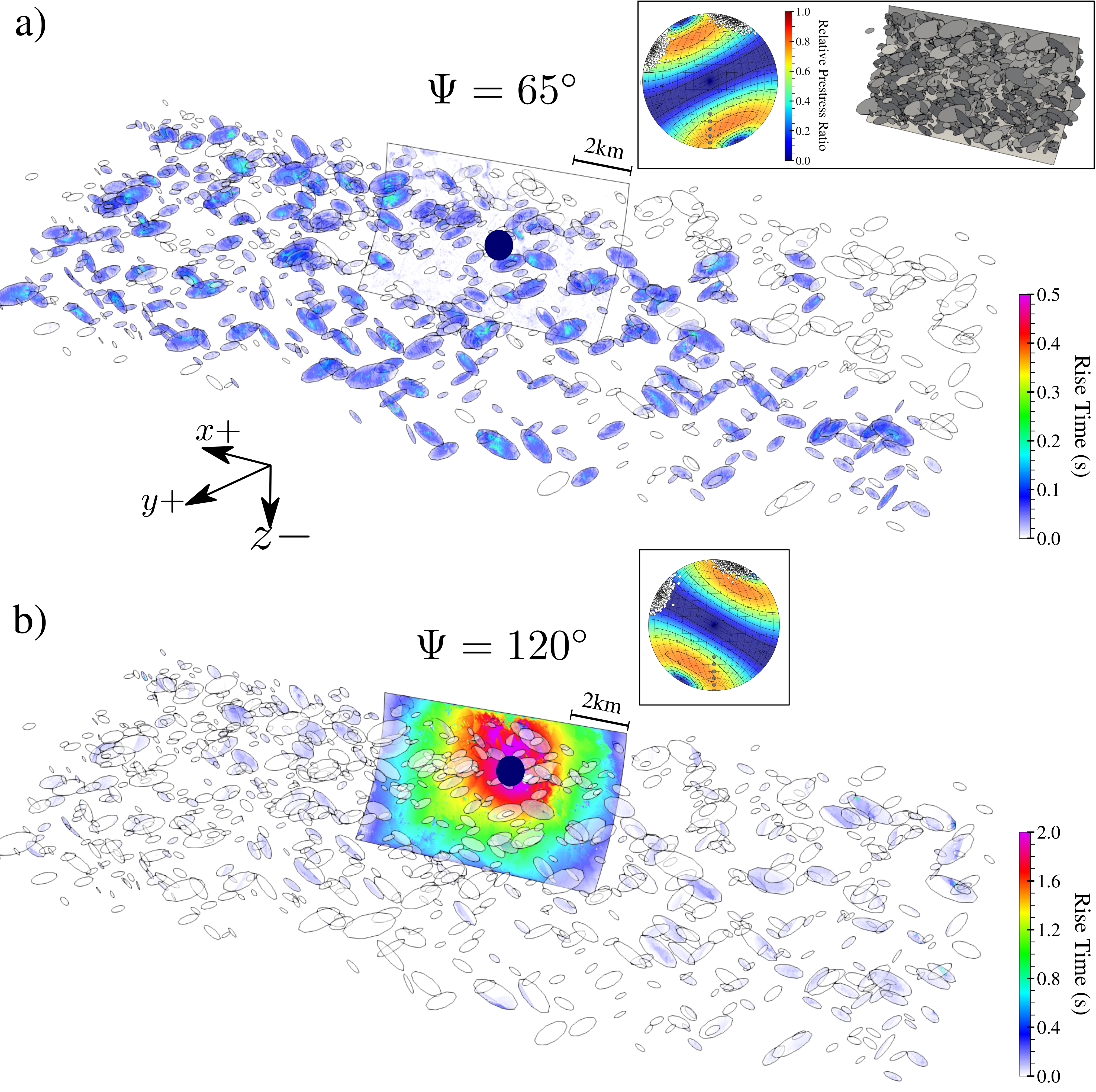}
    \caption{Rise time of two scenarios shown in an exploded view: (a) cascading earthquake ($\Psi=65^\circ$) with rupture occurring only on fractures, and (b) non-cascading earthquake ($\Psi=120^\circ$) with rupture occurring on the main fault and a subset of fractures. The inset in the top right panel displays the non-exploded model. The orientation-dependent pre-stress ratios are shown in lower-hemisphere projections, each fracture marked by a white circle. The black circle marks the hypocenter. It should be noted that the non-cascading earthquake has a longer rise time.}
    \label{fig:riseTime}
\end{figure}

\begin{figure}[htp!]
    \centering
    \includegraphics[width=0.95\textwidth]{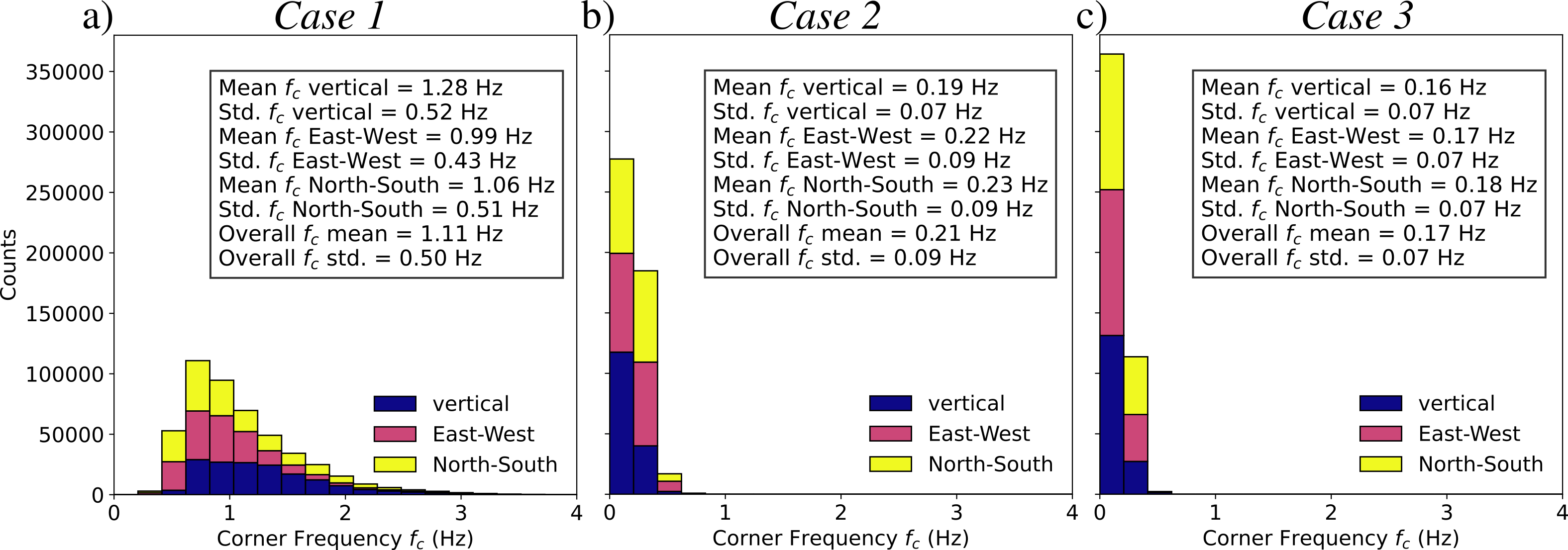}
    \caption{Statistics of equivalent near-field characteristic frequency $f_c$ for all surface stations in three components (vertical, East-West, and North-South). a) $f_c$ statistics for \textit{Case 1}. b) $f_c$ statistics for \textit{Case 2}. c) $f_c$ statistics for \textit{Case 3}. }
    \label{fig:fc_statistics}
\end{figure}

\begin{figure}[htp!]
    \centering
    \includegraphics[width=0.7\textwidth]{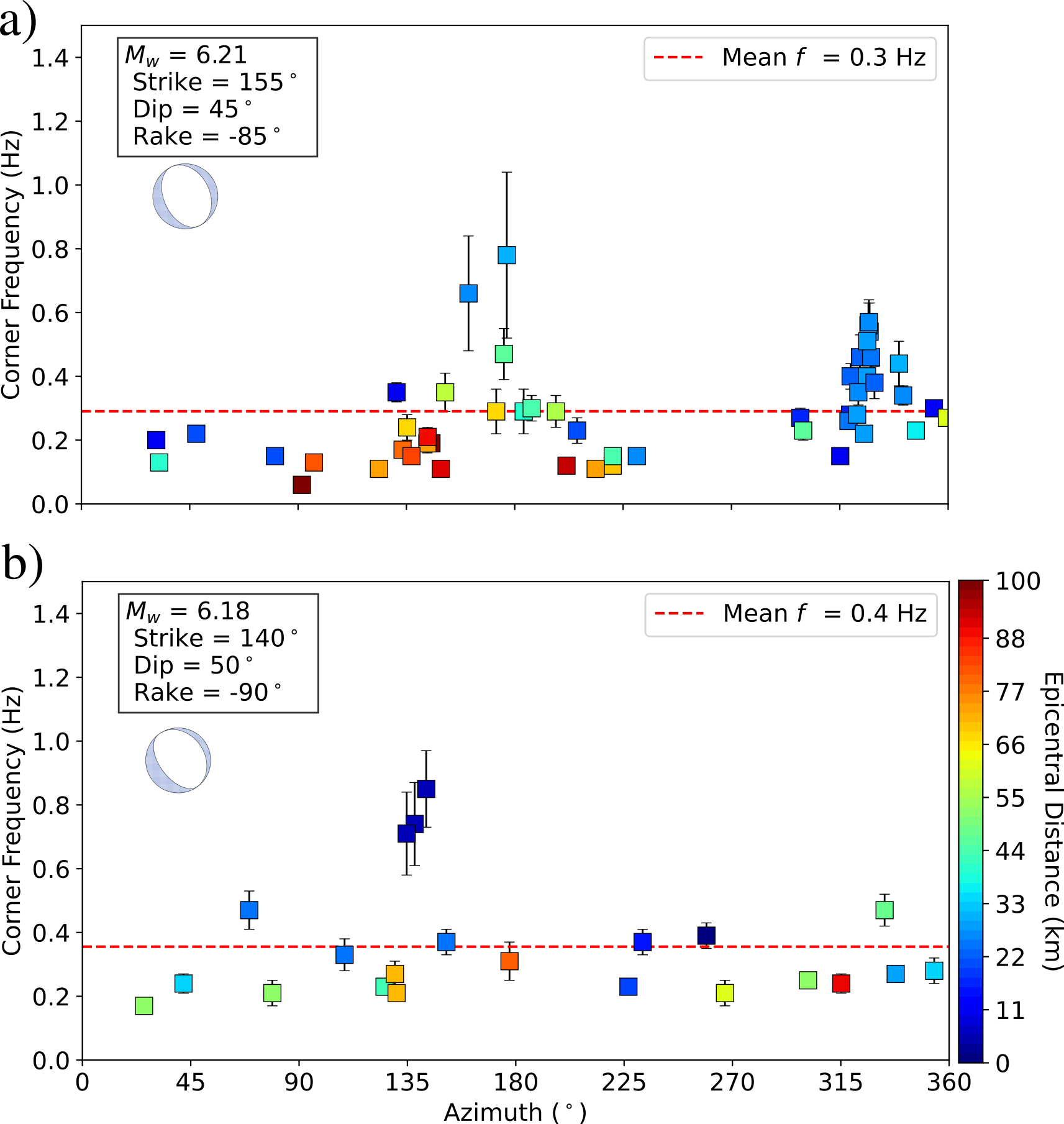}
    \caption{Corner frequency of recent well-recorded earthquakes with respect to azimuth relative to its epicenter for (a) the 2016 $M_\mathrm{w}$ 6.2 Central Italy and (b) the 2009 $M_\mathrm{w}$ 6.2 L'Aquila. The focal mechanism of each event is shown in the top left of each figure.}
    \label{fig:fc_realEarthquake}
\end{figure}

\begin{figure}[htp!]
    \centering
    \includegraphics[width=0.9\textwidth]{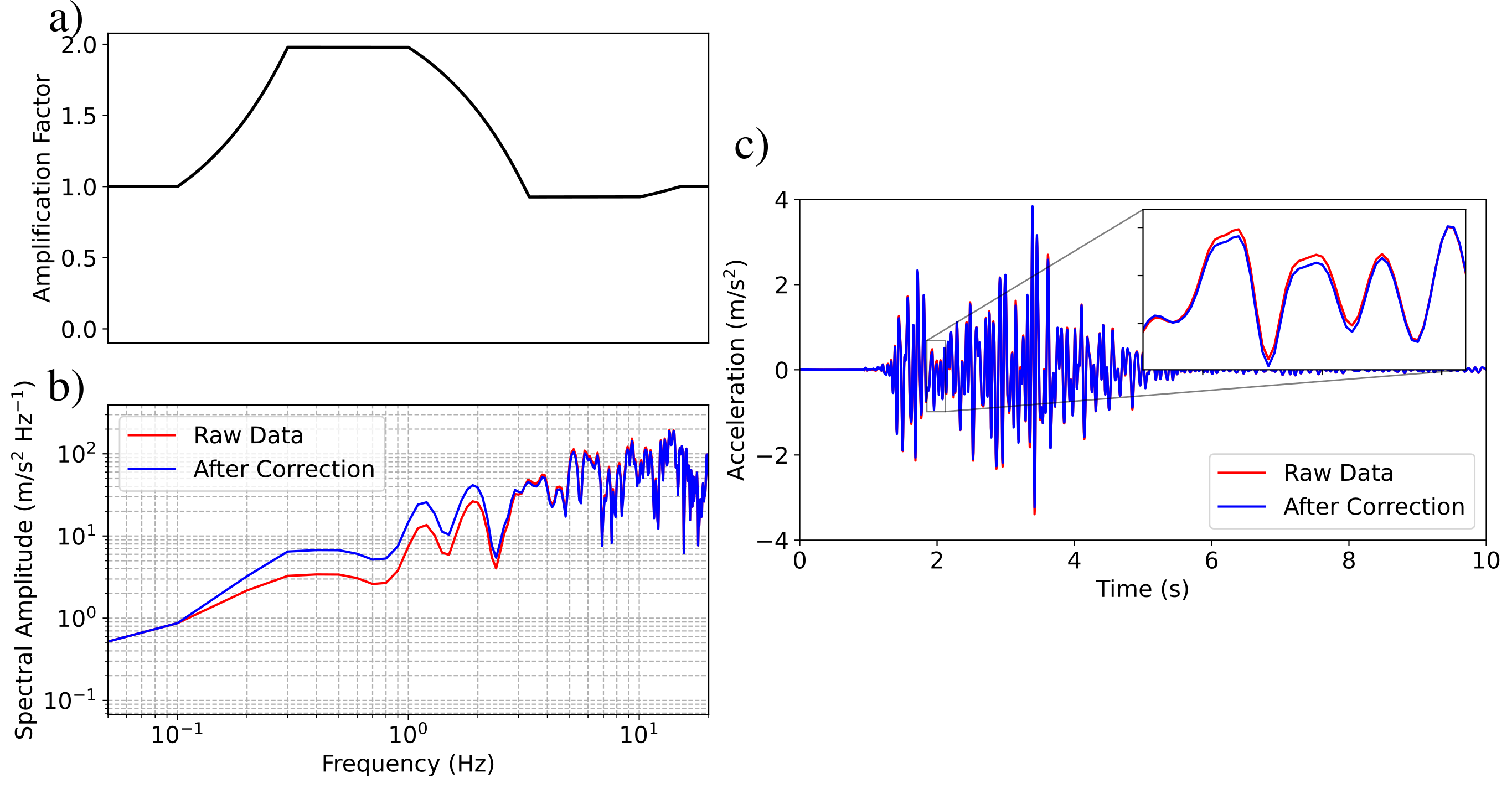}
    \caption{Site-amplification correction factor \cite{borcherdt1994estimates, borcherdt2002empirical} to correct for the homogeneous S-wave speed (3.464 km/s) in the simulations with respect to the reference value of $V_{S30}$ = 0.76 km/s used in the GMMs.
    This correction factor has been successfully applied in numerous ground-motion simulation studies (e.g., \citeA{mai2010hybrid, imperatori2012sensitivity, moczo2018key})}
    \label{fig:corrections}
\end{figure}

\begin{figure}[htp!]
    \centering
    \includegraphics[width=0.9\linewidth]{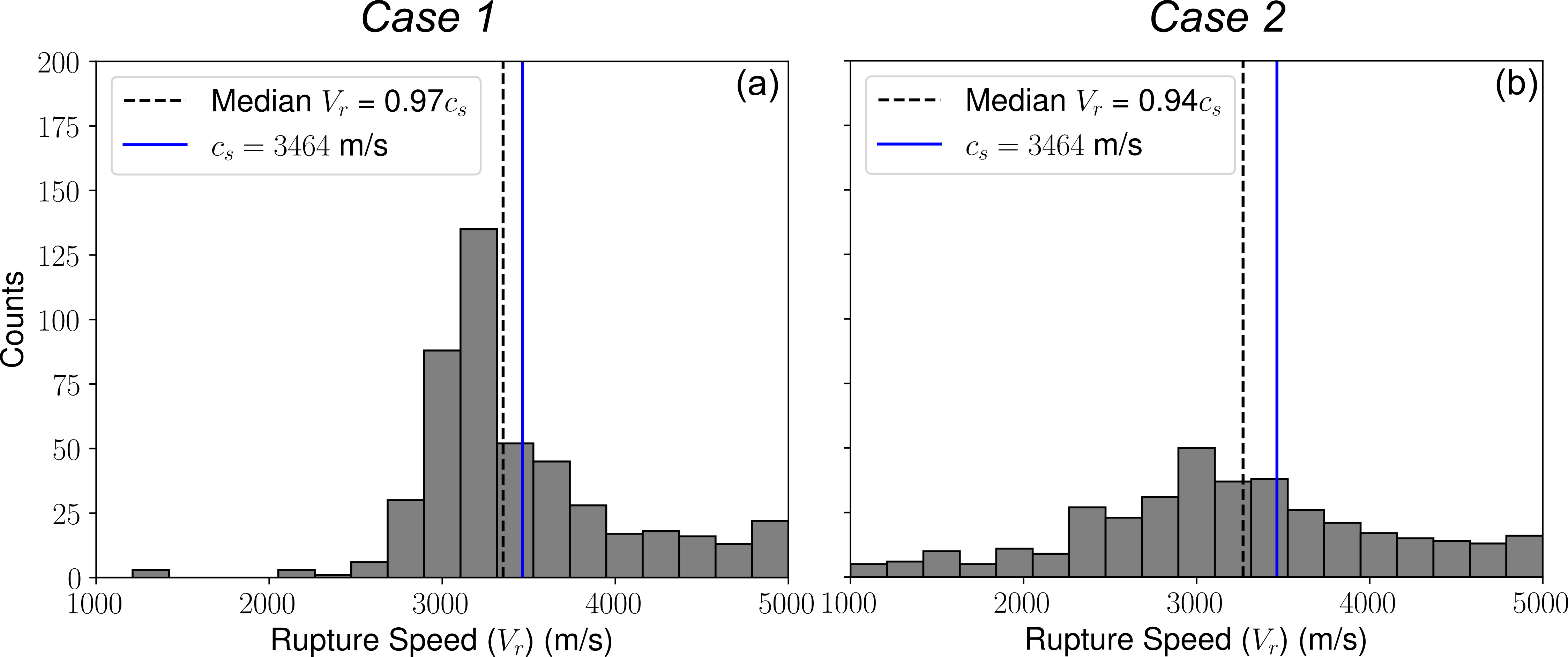}
    \caption{Histogram of the rupture speed for \textit{Cases 1} (cascading rupture within the fracture network) (a) and \textit{2} (rupture with off-fault fracture slip) (b). The black dashed line corresponds to the median of the rupture speed. The blue line illustrates the S-wave speed considered in this study.}
    \label{fig:rupture_speed_statistics}
\end{figure}

\end{document}